\documentclass[10pt,aps,prb, twocolumn,superscriptaddress,amsmath,amssymb]{revtex4-1}
\usepackage{graphicx} 
\usepackage{amsmath}
\usepackage{xcolor}
\usepackage{dcolumn}   
\usepackage{bm}        
\usepackage{verbatim}   
\usepackage{gensymb}
\usepackage{epstopdf}
\usepackage{footmisc}
\usepackage{natbib}
\usepackage{subcaption} 
\usepackage{multirow}
\usepackage{physics}
\usepackage{xfrac}
\usepackage{braket}
\usepackage{mathtools}
\usepackage{comment}
\usepackage{threeparttable}
\usepackage[referable]{threeparttablex}
\usepackage{xcolor}

\usepackage[colorlinks=true,linkcolor=blue,citecolor=blue,urlcolor=blue]{hyperref}

\usepackage{graphicx}
\usepackage{dcolumn}
\usepackage{bm}
\usepackage{soul}
\usepackage{color}
\usepackage{xcolor}
\usepackage{amsmath}
\usepackage{amssymb}
\usepackage{float}
\usepackage{natmove}
\usepackage{multirow}
\usepackage{setspace}
\usepackage{array}
\usepackage{booktabs}
\usepackage{rotating}
\usepackage{ulem}
\usepackage[colorlinks=true,linkcolor=blue,citecolor=blue,urlcolor=blue]{hyperref}
\usepackage{physics}
\usepackage{mathrsfs}
\usepackage{cprotect}
\usepackage{mathtools}
\usepackage[capitalise]{cleveref}
\usepackage[version=3]{mhchem} 
\definecolor{Grey}{rgb}{0.50,0.50,0.50}
\definecolor{Blu}{rgb}{0.00,0.00,1.00}
\definecolor{Red}{rgb}{1.00,0.00,0.00}
\definecolor{Green}{rgb}{0.20,0.50,0.20}
\definecolor{Magenta}{rgb}{0.60,0.00,0.60}
\definecolor{BluBondi}{rgb}{0.00,0.58,0.71}
\definecolor{Orange}{rgb}{0.95,0.46,0.17}
\definecolor{Red}{rgb}{1.00,0.00,0.00}

\newcommand{\editor}[2]{%
  \expandafter\newcommand\csname #1note\endcsname[1]{%
    \textcolor{#2}{{\it (\textbf{#1:} ##1)}}}%
  \expandafter\newcommand\csname #1\endcsname[1]{%
    \textcolor{#2}{##1}}%
  \expandafter\newcommand\csname #1cancel\endcsname[1]{%
    \textcolor{#2}{\sout{##1}}}%
  \expandafter\newcommand\csname #1change\endcsname[2]{%
    \textcolor{#2}{\sout{##1} ##2}}%
  \newenvironment{#1text}{\color{#2}}{\color{black}}
}

\editor{MB}{Green}
\editor{IO}{Magenta}
\editor{GP}{BluBondi}
\editor{PB}{cyan}

\newcommand{\drop}[1]{}

\renewcommand{\emph}{\textit}

\newcommand{\musr}{$\mu$SR}

\newcommand{\Mudot}{Mu$^{\mbox{$\boldsymbol{\cdot}$}}$}
\newcommand{\Hdot}{H$^{\mbox{$\boldsymbol{\cdot}$}}$}
\newcommand{\muO}{$\mu_{\mathrm{O}}$}
\newcommand{\muT}{$\mu_{\mathrm{T}}$}
\newcommand\aiidamuon{\texttt{aiida-muon}}

\begin{document}
\title{Automated computational workflows for muon spin spectroscopy}
\author{Ifeanyi J. Onuorah}
\thanks{IJO and MB contributed equally.}
\affiliation{Department of Physics and Earth Sciences, University of Parma, 43124 Parma, Italy}
\author{Miki Bonacci}
\thanks{IJO and MB contributed equally.}
\affiliation{Laboratory for Materials Simulations (LMS), Paul Scherrer Institut (PSI), CH-5232 Villigen PSI, Switzerland}
\author{Muhammad M. Isah}
\affiliation{Dipartimento di Fisica e Astronomia ``A. Righi'', Universit\'a di Bologna, I-40127 Bologna, Italy}
\author{Marcello Mazzani}
\affiliation{Department of Physics and Earth Sciences, University of Parma, 43124 Parma, Italy}
\author{Roberto De Renzi}
\affiliation{Department of Physics and Earth Sciences, University of Parma, 43124 Parma, Italy}
\author{Giovanni Pizzi}
\email{giovanni.pizzi@psi.ch}
\affiliation{Laboratory for Materials Simulations (LMS), Paul Scherrer Institut (PSI), CH-5232 Villigen PSI, Switzerland}
\author{Pietro Bonfà}
\email{pietro.bonfa@unipr.it}
\affiliation{Department of Physics and Earth Sciences, University of Parma, 43124 Parma, Italy}
\date{\today}
\begin{abstract}

Positive muon spin rotation and relaxation spectroscopy is a well established experimental technique for studying materials. It provides a local probe that generally complements scattering techniques in the study of magnetic systems and represents a valuable alternative for materials that display strong incoherent scattering or neutron absorption. 
Computational methods can effectively quantify the microscopic interactions underlying the experimentally observed signal, thus substantially boosting the predictive power of this technique. Here, we present an efficient set of algorithms and workflows devoted to the automation of this task. In particular, we adopt the so-called DFT+$\mu$ procedure, where the system is characterised in the density functional theory (DFT) framework with the muon modeled as a hydrogen impurity. We devise an automated strategy to obtain candidate muon stopping sites, their dipolar interaction with the nuclei, and hyperfine interactions with the electronic ground state. We validate the implementation on well-studied compounds, showing the effectiveness of our protocol in terms of accuracy and simplicity of use.
\end{abstract}
\maketitle

\section{Introduction}
Positive muon spin resonance spectroscopy (\musr) is a powerful experimental probe to investigate the properties of a wide range of condensed matter systems at the atomic scale, through the interaction between the muon spin and the atomic environment in its vicinity~\cite{yaouanc2011,blundell2022}. In analogy to nuclear magnetic resonance (NMR), \musr{} is a highly sensitive local probe of ordered magnetism, including very weak magnetic moments, thanks to the large gyromagnetic ratio $\gamma_{\mu}/(2\pi) \approx$ 135.5 MHz/T of spin $I=\frac 1 2$ implanted muons~\cite{schenck1995}. It is mostly used to study the magnetic properties of materials, where the muon acts as a tiny magnetometer at an interstitial position in the sample, but it also provides an effective tool to study superconducting order parameters, light particle diffusion, and charge related phenomena~\cite{blundell2022}.  
The experimentally acquired signal is the muon spin polarization function, which is the time evolution of the muon spin projected along a given direction.
In the conventional scheme for data analysis, the main features of the signal are determined by a best fit to an effective model that contains one or more Larmor precession frequencies and their decay in time, or simply exponential decays that result from a distribution of local fields at the muon sites of either nuclear or electronic origin.
This approach conveys important information of the system, such as order parameters, transition temperatures, presence of phase separation, but generally delivers only limited microscopic information on the sample under investigation.
This is because the muon stopping site and its interactions with the host are generally unknown (at variance with NMR where the position of the nucleus is precisely known). 
A few experimental approaches such as the measurement of the anisotropy of the Knight/paramagnetic shifts, or of level crossing resonances, can be used to find muon stopping sites~\cite{kiefl1987,schultz2005,schenck1995}, but they require both extensive beam time and large single crystal compounds, which are not generally available.
On the contrary, computational approaches based on the Density Functional Theory (DFT), dubbed DFT+$\mu$ in the literature, have been recently shown to be less costly and very accurate in finding the muon stopping site and its hyperfine interactions\cite{moeller2013,bonfa2013, bonfa2016,bonfa2015,onuorah2018,onuorah2019,blundell2023}. 
Furthermore, in some cases DFT+$\mu$ can be used to describe the long-range structure of a ground-state magnetic order~\cite{PhysRevB.97.224508,lamura2020, bonfa2021} or to study the extent of the muon-induced perturbation, which generally does not affect the intrinsic properties of the compound being studied, with a few notable exceptions readily identified with the ab-initio approach~\cite{foronda2015,dehn2020, huddart2021}. 
 
Nowadays, the DFT+$\mu$ method is well established and its adoption for the interpretation of \musr{} data is becoming more frequent. However, it is still challenging for non-experts to setup the machinery required to perform the simulations, acquire sufficient expertise to use the instrument efficiently, and afford the considerable human intervention required to track and coordinate the large number of calculations involved. 
These issues have limited a routine usage of the ab-initio approaches, as well as its adoption in on-line data analysis alongside experiments. To overcome these challenges, an automated framework is highly desirable.

At the same time, the tremendous increase in the high performance computational (HPC) resources has led to the development of automation infrastructures that allow researchers to define workflows, manage calculations and effectively store and organize results into databases.
A number of workflows have already been developed to carryout several high-throughput (HT) studies aimed at material design and discovery targeting different properties or applications~\cite{mounet2018,doi:10.1021/acs.jpcc.2c06733,Torelli_2019,PhysRevMaterials.1.034404,chen2019,Zhang_2021, torelli2020,C9NA00588A, doi:10.34133/2022/9857631,matprojvalues2020, horton2019}. Automated computational approaches are also a valuable tool for the interpretation of other spectroscopic measurements, such as X-ray photoelectron spectroscopy~\cite{xrayph2022} and X-ray absorption spectroscopy~\cite{Zheng2018,meng2024,Guo2023}, NMR chemical shifts~\cite{nmrhtp2,nmrhtp}, electric field gradients (EFG) for nuclear quadrupole resonance (NQR) \cite{nqrhtp} and IR and Raman spectra~\cite{ramanhtp,inframan2}.

In this paper, we present robust and fully automated workflows implementing protocols to find the muon stopping sites and quantifying its interactions with the hosting sample. The workflow takes advantage of the AiiDA (Automated Interactive Infrastructure and Database for Computational Science) infrastructure~\cite{AiiDA2016,AiiDA2020,AiiDA2021}, which is designed to automate, manage, and store computational models making them easily shareable with the scientific community. 
A few other tools for muon site identification have been published in recent years (see for example the MuFinder Python package~\cite{mufinder2022} or the MuonGalaxy platform \cite{materialsgalaxy}). Our approach differs in at least three aspects. First, by leveraging on the AiiDA platform, we can incrementally adopt relevant methods and tools from other application domains: for example, accurate and automated treatment of self-interaction correction errors in strongly correlated systems~\cite{TIMROV2022108455, Bastonero_aiidahp}, or general purpose graphical user interfaces (GUIs) such as AiiDAlab \cite{AiiDAlab}. In addition, the platform allows the design of ``code agnostic'' workflows which can work with any method that provides forces and total energies for a given structure~\cite{Huber_2021,Bosoni_2024}. Second, we provide complete information on the interaction between the muon and its neighbouring environment, considering both hyperfine parameters and the dipolar interaction with the nuclear moments. We note, however, that DFT+$\mu$ must be compared to $T=0$ \musr{} results and that muon thermally activated motion, important for the interpretation of many experiments, is beyond the scope of  the present workflow.
Third, thanks to AiiDA provenance model, all workflows and their provenance are fully tracked and made reproducible and easy to share with the rest of the community.

The paper is organized as follows: we first describe the DFT+$\mu$ approach. Then, in the ``Results and discussion'' section, we first describe the algorithms used to automate the calculations and then illustrate two relevant aspects of the workflows with examples: the novel efficient method developed for finding the converged supercell size (that we validate on neutral and charged calculations in a metal, LaCoPO, and an insulator, LiF) and the crucial importance of the inclusion and automated treatment of Hubbard U parameters (in CuO).
Finally, the whole workflow is used to study the muon localization and interactions in selected compounds, validating it against experiments. The systems we consider include the Kagome-structured superconductors (AV$_3$Sb$_5$ A= K, Rb, Cs), a fluoride (CaF$_2$, where the strong $^{19}$F--muon dipolar interaction provides a direct experimental benchmark), an antiferromagnetic insulator (La$_2$NiO$_4$) and a ferromagnetic metal (LaCoPO). These examples also help us illustrate the workflow features. Finally, future development directions are discussed.

\section{DFT+\texorpdfstring{$\mu$}{Lg} and muon interactions}
\label{sec:bkg}
Within the DFT+$\mu$ approach, the muon implantation site is identified by performing DFT calculations in the Born-Oppenheimer approximation, where the positive muon is treated as an extremely diluted interstitial hydrogenoid impurity. A plane-wave basis set is often used for crystalline compounds.
This is an effective choice for the description of the host system but requires the introduction of supercells in order to suppress the spurious interactions between the muon and its periodic replica. 

Experimentally, two different charge states are distinguished: the diamagnetic and the paramagnetic muon. These correspond to the two limiting cases of a bare muon or a muon strongly bound to an electron.
These two states are accordingly simulated by considering a positive impurity, Mu$^+$, for the diamagnetic state and neutral impurity, \Mudot~for the paramagnetic one (in analogy with H$^+$ and \Hdot, where the dot is the symbol for an unpaired radical state) obtained within the DFT approach with a charged or neutral supercell, respectively\footnote{The overall required charge neutrality is ensured by a compensating background.}. The starting charge is evolved with the standard self-consistent scheme and the resulting muon state is inspected from the converged charge density.
Notably, in general, negligible differences are observed in metals \footnote{On the hosting system, the addition of a hydrogen atom generally produces an effective electronic doping, that is progressively reduced by increasing the supercell dimensions.} due to the effective screening of the positive charge by the conduction electrons \cite{moeller2013,blundell2023}, whereas distinct solutions may be obtained in insulators.

The identification of candidate interstitial sites for the muon is performed by sampling the voids of the lattice structure. A regular grid or a set of random points is generated. The set of initial positions is later reduced by considering the lattice symmetry of the crystal hosting the muon.
For each initial position, all atomic positions are optimized to accommodate the muon until forces acting on all the atoms vanish below a given threshold. A minor technical point is that residual symmetries after the insertion of the muon should be discarded and the cell parameters all remain fixed to avoid introducing spurious stresses due to the (infinitely diluted) impurity \cite{DefectsInSemiconductors}. The procedure results in optimized muon positions that correspond to distinct local minima in the potential energy surface, usually identified as \emph{candidate muon sites}. Each unique crystallographic position is characterized by a total energy and, although muons might stop at more than one site, the lowest energy ones are assumed to correspond to the most probable stopping sites in the experiment.  

The muon mass being one ninth of the hydrogen mass, its zero point energy (ZPE), typically a fraction of an electronvolt, is not negligible. This implies that, even at $T=0$, it is important to assess whether higher energy candidate sites are metastable by virtue of the muon zero point motion (ZPM). A self-consistent treatment of the muon ZPM is not currently included, but is planned in future code developments (see ``Future developments'' section). 

In order to verify the correctness of the identified muon site, it is standard practice to compare the predicted spin polarization function with the one obtained in well characterized experiments. In the following paragraphs we briefly describe how this is generally done.

\subsection{Experimental benchmarks}
\label{sec:experiments}
 
In a zero-field \musr{} experiment, the muon spin polarization is dictated by interactions with the nuclear and electron magnetic moments. 
Randomly directed neighbor spins (both nuclear and electronic) are well approximated by an isotropic Gaussian distribution of static dipolar fields with zero average and distribution width $\Delta B$, giving rise to the Gaussian Kubo-Toyabe polarization function~\cite{Blundell_book_2021}
\begin{equation}
\label{eq:KT}
    P(t) = \frac{2}{3} \left[ 1 - \Delta^2 t^2\right]   e^{-\Delta^2 t^2 / 2} + \frac{1}{3}
\end{equation}
where $t$ is time and $\Delta=\gamma_\mu\Delta B$ is the decay rate. This function may fit fast muon spin depolarization in highly disordered magnets, and the much slower spin relaxation of non magnetic materials with a static neighborhood of the muon including large nuclear moments. Incipient dynamics may be monitored by strong collision modification of Eq.~\eqref{eq:KT}.

Non-magnetic fluorides (e.g. LiF, CaF) deserve a special mention as the prototypical example where a strong dipolar interaction among the muon magnetic moment and few close nearest-neighbor large nuclear moments  provides a unique signature of the muon stopping site. \cite{PhysRevB.33.7813,bonfa2013,PhysRevB.87.121108,moeller2013} A frequent geometry is that of a straight symmetric F--Mu$^+$--F bond, whose distances are encoded in the classical dipolar interaction ${\cal H}_{\mathrm{dip}}= -\mathbf m_\mu\cdot \mathbf B_{\mathrm{dip}}$, where the dipolar field of the $N=2$ fluorine nuclear moments is written as 
\begin{equation}
\mathbf{B}_{\mathrm{dip}}=\frac{\mu_0}{4\pi}\sum_{i=1}^N{\left[\frac{3 \mathbf{r}_i (\mathbf{m}_i \cdot \mathbf{r}_i)} {r_i^5}    - \frac{\mathbf{m}_i } {r_i^3} \right]},
\label{eq:muobdipfinal}
\end{equation}
$\mathbf{m}_\mu, \mathbf{m}_i$ are the muon and fluorine magnetic moments, and $r_i$ is the distance between the $i^{\text{th}}$ nucleus and the muon site.

The interaction with electrons in ordered magnets is instead obtained with the usual minimal substitution $p \rightarrow p + A_\mu$, where $A_\mu$ is the magnetic vector potential associated with a muon magnetic moment \cite{Blundell_book_2021}.
Straightforward elaborations show that all contributions are linear in the muon magnetic moment $\mathbf m_\mu$ and can therefore be expressed as ${\cal H}^\prime= -\mathbf m_\mu\cdot \mathbf B_\mu$, where  $\mathbf B_\mu$ is an effective \emph{internal} magnetic field.
The internal field is conventionally separated into
\begin{equation}
\mathbf{B}_\mu = \mathbf{B}_{\mathrm{dip}} + \mathbf{B}_{\mathrm{c},}
\label{eq:bdipform1}
\end{equation}
where $\mathbf{B}_ {\mathrm{dip}}$ is the dipolar term and $\mathbf B_{\mathrm{c}}$ is the contact term, that are straightforwardly defined in the classical interaction \cite{Jackson}. 
The former is generally well reproduced by the classical approximation given by Eq.~\eqref{eq:muobdipfinal}, assuming that distant electronic magnetic moments are point-like classical dipoles localized on magnetic ions. This classical sum is generally performed with the Lorentz sphere approach~\cite{muesr2018}. 
From Eq.~\eqref{eq:muobdipfinal}, it is evident that the crucial parameters that drive this contribution is the magnetic moment $\mathbf
{m}_i$ and the distance between the magnetic atoms and the muon. The latter is often affected by the short ranged perturbation of the host lattice induced by the implanted muon \cite{mufinder2022}.
$\mathbf{B}_c$ is the Fermi contact term, which instead requires a quantum-mechanical treatment.
A good estimate is obtained from DFT calculations~\cite{onuorah2018}, and  it is given by
\begin{align}
\bm{B}_{\text{c}} & = \frac{2\mu_0\mu_{B}}{3}\rho_{s}(\bm{r}_{\mu}) \bm{\hat{z}} \label{eqn2:contact_hyperfine_field}
\end{align}
where  $\mu_0$ is the vacuum permeability, $\mu_{B}$  is the Bohr magneton and $\rho_s =[\rho^{\uparrow}_s - \rho^{\downarrow}_s ]$ is the spin density at the muon position $\mathbf{r}_\mu$.  $\rho^{\sigma}_s =  \left| \psi^{\sigma} \right|^2$ is the density associated to each spin component $\sigma = \uparrow,\downarrow$ along the direction  $\bm{\hat{z}}$  of the magnetic moment at the muon,
which coincides with the bulk magnetization direction and it may be determined by integration inside a suitable sphere in the general case.

\begin{figure*}[tb]
\centering
\includegraphics[width=14cm]{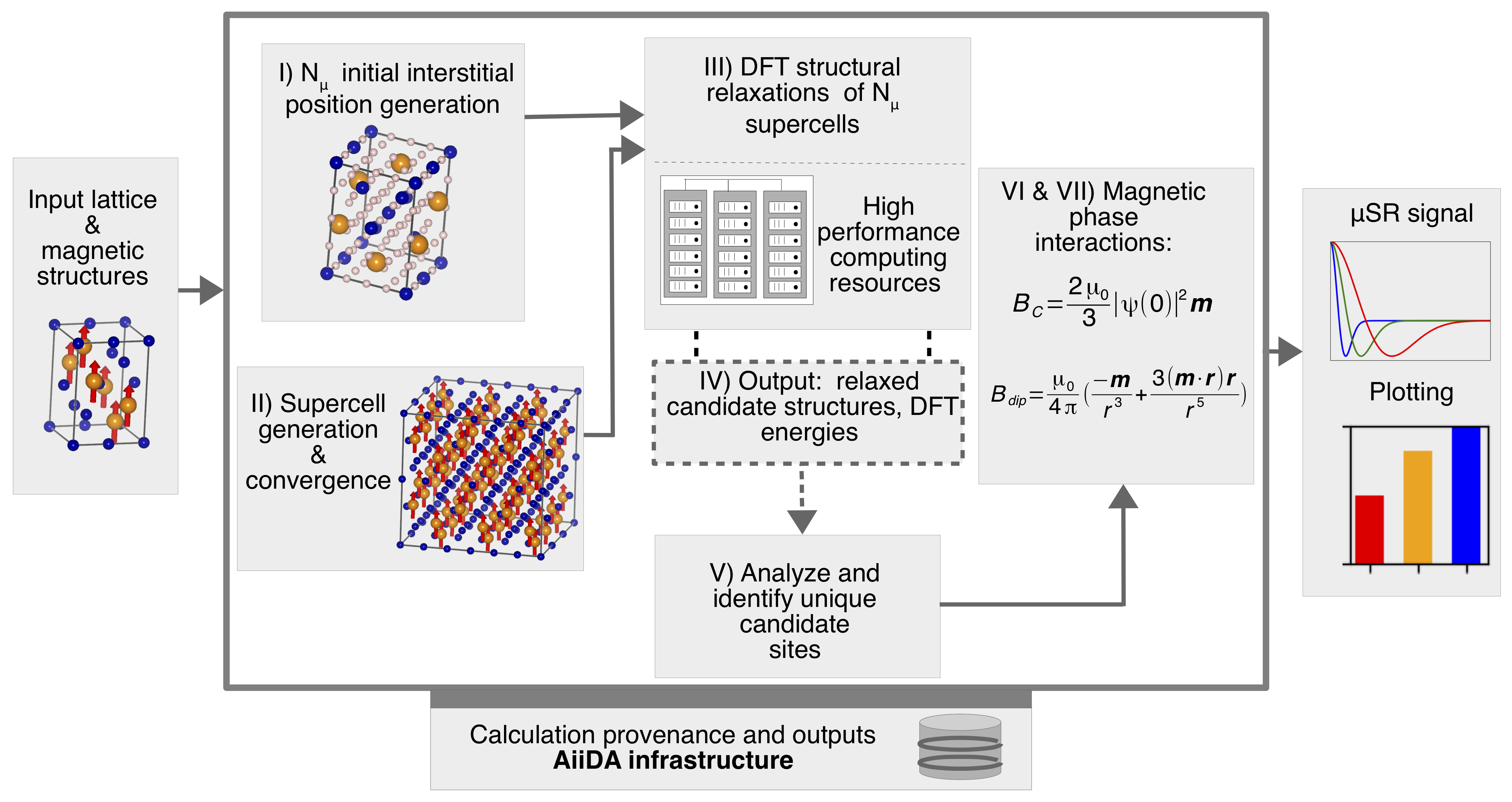}
\caption{Schematic representation of the \aiidamuon{} workflow. For a description of each box, see points I-VII in section ``Automated workflow design'' in the main text.}
\label{fig:workflow}
\end{figure*}

\section{Results and discussion}

\subsection{Automated workflow design}
\label{sec:automation}
The computational approach described above is automated in a Python workflow implemented within the AiiDA framework. The workflow is packaged in the \aiidamuon{} plugin (see ``Code availability'' section), which contains the \verb|FindMuonWorkchain| (AiiDA workflow), the top-level workflow that we expose to users (and internally calls other workflows or calculations).
The workflow features include handling of crystal and magnetic structures leveraging on the Atomic Simulation Environment (ASE)~\cite{HjorthLarsen_2017} and Python Materials Genomics (Pymatgen)~\cite{ONG2013314} libraries, generation of the initial interstitial muon positions, and automatic convergence of the supercell size.  
The AiiDA platform handles the interaction with HPC facilities and performs post-processing operations to fetch and parse results, storing them in a database and thus providing efficient query capabilities.
As already mentioned, another important feature provided by AiiDA is the definition of ``code agnostic'' workflows, thus potentially providing  different levels of accuracy/speed for the DFT+$\mu$ problem. However, at the moment the workflow only interfaces with the Quantum ESPRESSO plane-wave DFT code~\cite{qe2009} via the \verb|aiida-quantumespresso| plugin \cite{AiiDA2021}.

The schematic representation of the workflow is shown in Fig.~\ref{fig:workflow}. The input includes the crystallographic structure and, possibly, a magnetic order that can be provided, for instance, via files in CIF and mCIF format. Optional input settings for the DFT calculations are provided as Python variables and transformed into AiiDA data formats. The execution of the workflow involves seven steps, labelled I--VII and discussed below.

\begin{itemize}
\item[I)] Generate a number $N_\mu$ of initial muon interstitial sites.
\item[II)] Execute  the supercell (SC) convergence sub-workflow (see ``Supercell convergence workflow'' section), unless a given SC size is explicitly provided as input. 
\item[III)] Execute the structural relaxation of the $N_\mu$ supercells, typically in parallel, on HPC clusters.   
\item[IV)] Inspect and ensure that at least 60\% of the simulations of step III are completed successfully; if not, the workflow stops due to structural convergence failure.  
\item[V)] Collect the relaxed structures and their total energies, and cluster distinct stable structures on the basis of symmetry and total energy differences (see ``Unique sites selection'' section).

\item[VI)] If a magnetic order is provided as input, obtain the contact contribution ($B_{\mathrm{c}}$) to the local field from the spin-resolved electronic density computed with a dense reciprocal space grid for the distinct stable structures.

\item[VII)] Compute the muon dipolar field interactions at the relaxed structures and the input magnetic configuration using the MUESR code~\cite{muesr2018} interfaced as an AiiDA calculation function (\verb|calcfunction|). 
\end{itemize}

 Notably, thanks to the fault tolerant and fault resilient algorithms of the a\verb|iida-quantumespresso| \verb|PwBaseWorkChain|, the workflow in step III  can handle a range of typical errors, such as unconverged SCF calculations or hitting of walltime limits, ensuring that in most cases the calculations finish successfully. In step V, for magnetic compounds, it can happen  that crystallographically equivalent replica of a candidate muon site  may  be magnetically inequivalent \cite{bonfa2024}. When this happens, step III is reactivated, so that relaxed structures of missing magnetically inequivalent sites are obtained and added to the list. Calculations for the charged supercell for the the Mu$^+$ state (default) and neutral for the \Mudot~state (optional) are run independently and controlled in the workflow by the Boolean input parameter \verb|charged_supercell|.

Two technical points deserve discussion. First, lattice distortions introduced by the muon are always accounted for in the evaluation of dipolar and contact field contributions. Second, the computation of the dipolar contribution in step VII is performed using the magnetic structure provided as input, typically the experimental one, while the contact contribution in step VI is obtained from the ab initio description of the same system. This implies that the magnetic moments, or even the magnetic order obtained at convergence, may be different from the one provided as input. An example is illustrated in the LaCoPO section below.

The output of the workflow include: all the relaxed supercell structures and their total energies, the relaxed supercell structures for symmetry/magnetically inequivalent sites, their relative energies with respect to the lowest energy one, and the computed contact hyperfine and dipolar fields (if applicable). This output is stored in the database, with unique identifiers and appropriate metadata for future reference. AiiDA also allows users to export calculation files for ad-hoc post-processing.

Before concluding this section, two aspects of the automated workflow are further discussed.
We search for the candidate muon implantation site by sampling the interstitial space with a grid of $n_a\times n_b \times n_c$ regularly spaced initial positions, with the condition that each starting point is at least 1 \AA~ away from any host atom (if a point is closer to any atom, it is discarded). The three integers are determined as $n_i = \lceil a_i/d_{\mu} \rceil$ from the three lattice parameter $a_i = a,b,c$ and a unique input spacing $d_{\mu}$ (default value: 1 \AA). The lattice symmetry is finally used to eliminate equivalent points, generating a reduced total number  $N_\mu$ of initial muon interstitial sites.
The second point concerns calculations involving complex magnetism with non-collinear and/or incommensurate magnetic orders, that are still challenging for DFT-based automated workflows \cite{horton2019,Zhang_2021}. The description of non-collinear magnetism is indeed resource and time intensive as the spins have more degrees of freedom, thus making self-consistent convergence harder to achieve and often requiring the introduction of energy penalties to facilitate the minimization. Thus, in the current version we have restricted the DFT-based calculations to the collinear spin formalism only, with quantization axis along the $z$ direction, neglecting
spin-orbit coupling (and thus any anisotropy term). For this reason, magnetic contributions from $f$-electrons, where spin-orbit is important, cannot be currently treated correctly. However, non-collinear input array descriptions of the magnetic order are accepted for an accurate evaluation of the dipolar field sums. The DFT calculations are then performed on a collinear structure automatically obtained by projecting each moment along a global spin axis. The accuracy of this approximation varies case by case and the results obtained for non-collinear or incommensurate magnetic orders should therefore be carefully inspected by the user.

\subsection{Algorithms}
In the following we describe the main AiiDA calculation functions implemented in the workflow.
\subsubsection{Supercell convergence workflow}
\label{sec:supconv}

\begin{figure}[tb]
\centering
\includegraphics[width=8cm]{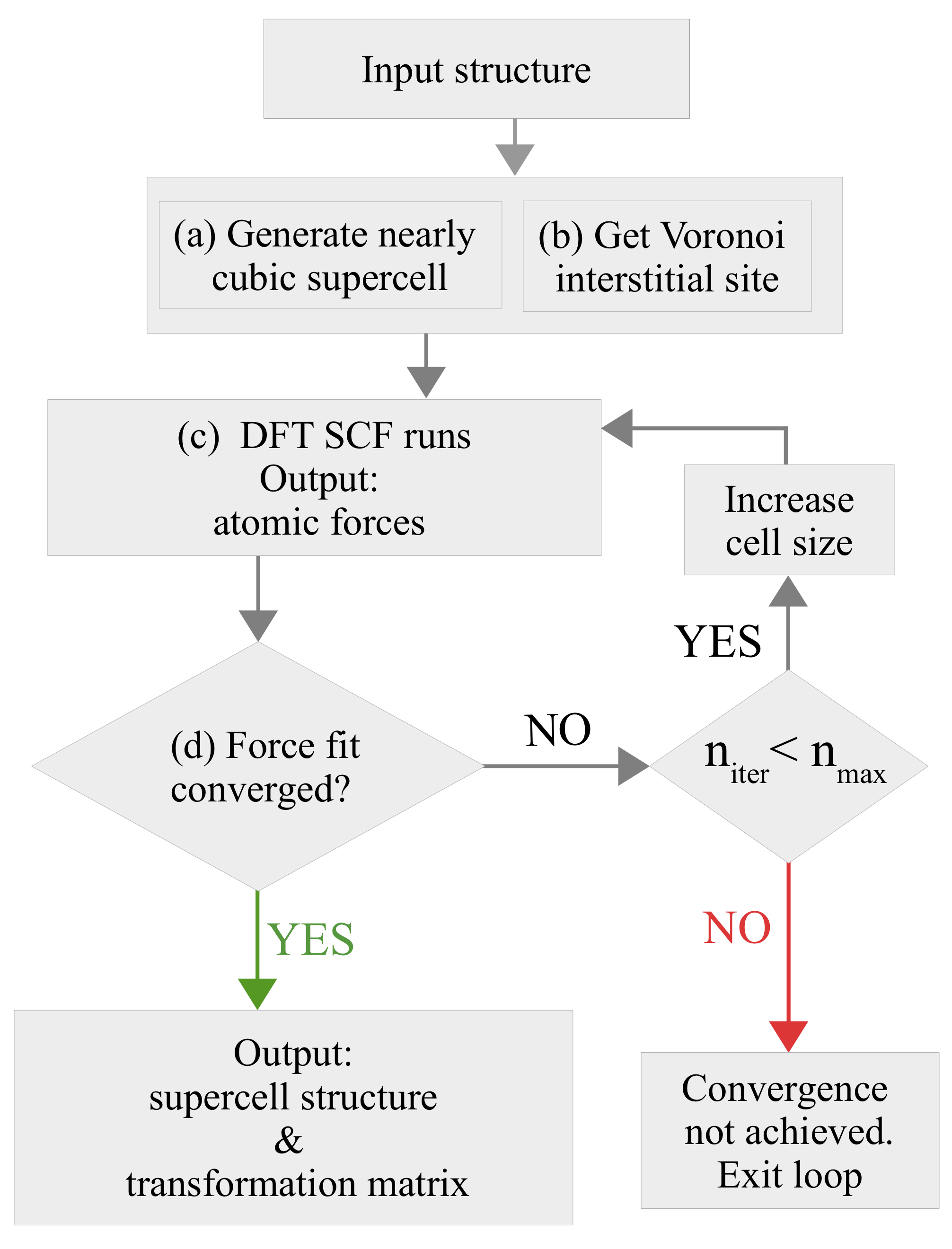}
\caption{The flowchart of the supercell convergence workflow implemented in the \texttt{IsolatedImpurityWorkChain}.}
\label{fig:conv}
\end{figure}

The minimal size of the supercell that allows the convergence of the calculations is a crucial parameter that must be automatically determined in the workflow.
The problem is well known~\cite{PhysRevB.58.1318,RevModPhys.86.253} and it has been extensively discussed in the context of defects and impurities in semiconductors. 
In the context of $\mu$SR, the convergence against the supercell size is obtained by considering different quantities of interest such as total energy, electric field gradients, muon hyperfine fields\cite{onuorah2018} or muon induced lattice distortions~\cite{PhysRevB.87.121108}.
Several of these quantities are extracted after structural optimization, a task that is computationally demanding.
Here we employ a less expensive technique, where the criterion for determining the supercell size is the convergence of the forces exerted by the impurity on the host atoms, which require a single self-consistent simulation. The new approach has been automated in a stand-alone workflow and packaged in the \texttt{aiida-impuritysupercellconv} plugin which contains the \texttt{IsolatedImpurityWorkChain}. It is also automatically used in the \verb|FindMuonWorkchain| if the input supercell matrix required for the muon calculations is not provided. 

\begin{figure}[!ht]
\centering
\begin{subfigure}{4.2cm}
\includegraphics[width=1.0\linewidth]{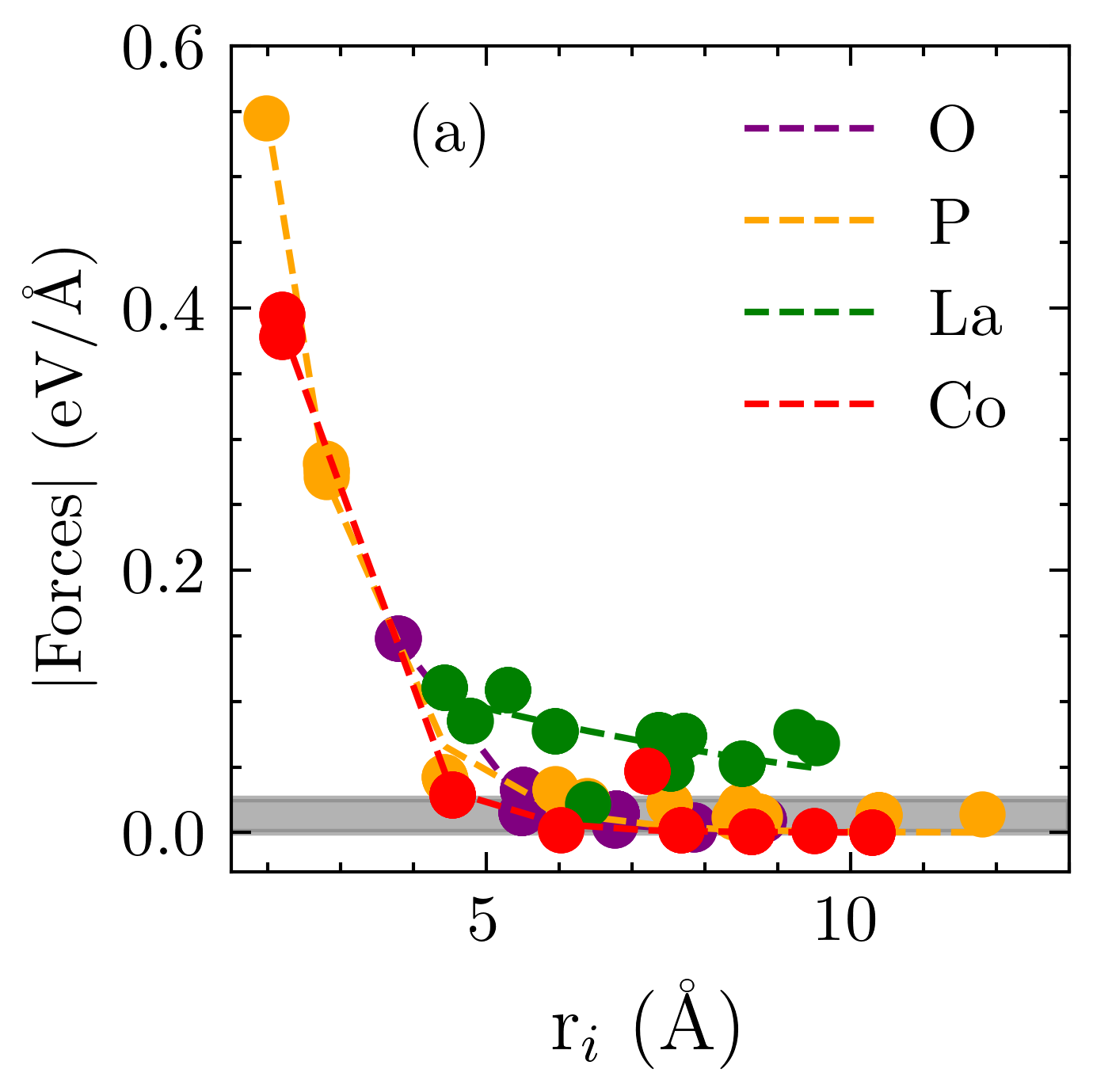}
  \label{fig:lcp2}
\end{subfigure}
\begin{subfigure}{4.2cm}
\includegraphics[width=1.0\linewidth]{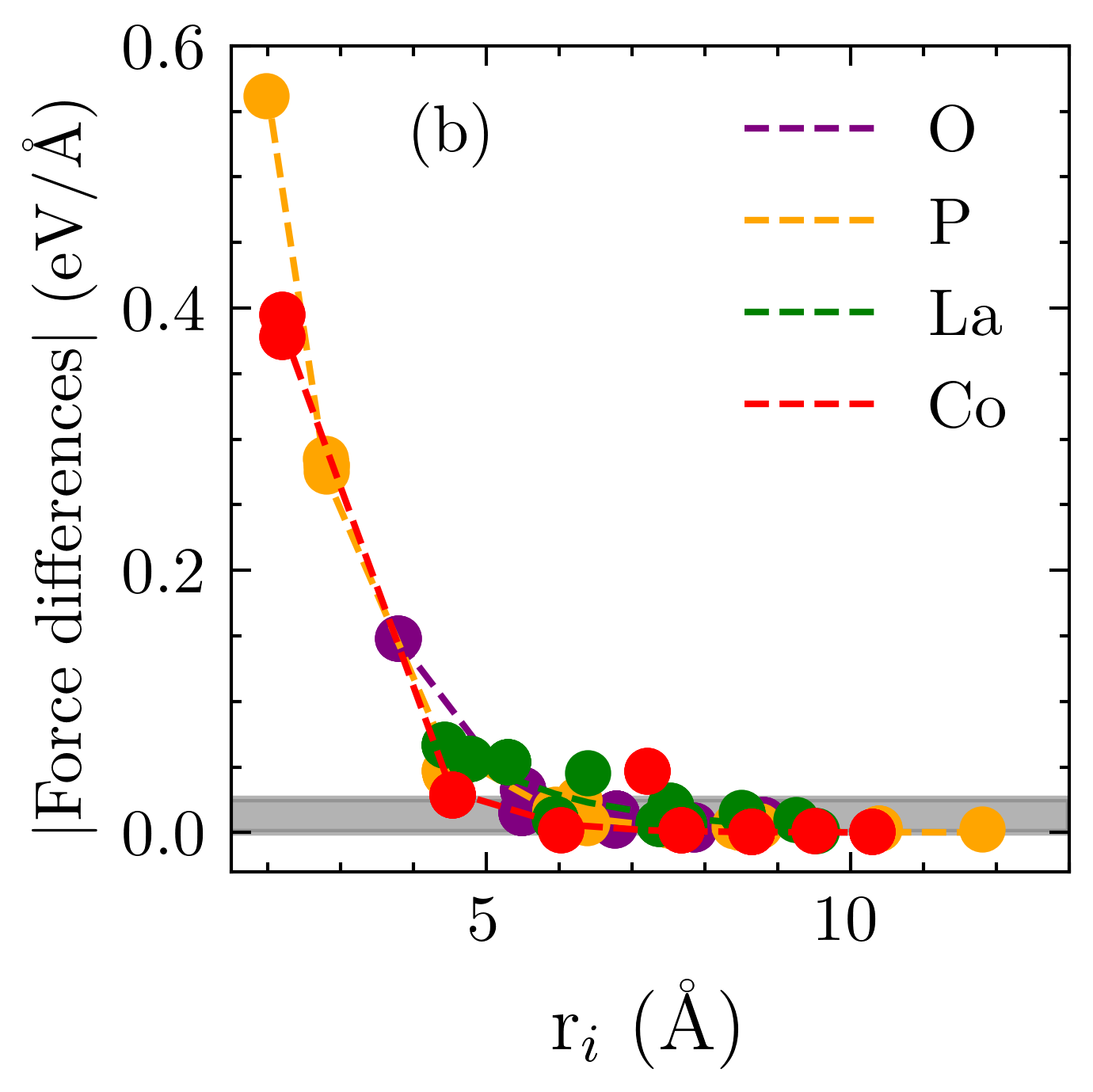}
  \label{fig:lcp22}
\end{subfigure}
\begin{subfigure}{4.2cm}
\includegraphics[width=1.0\linewidth]{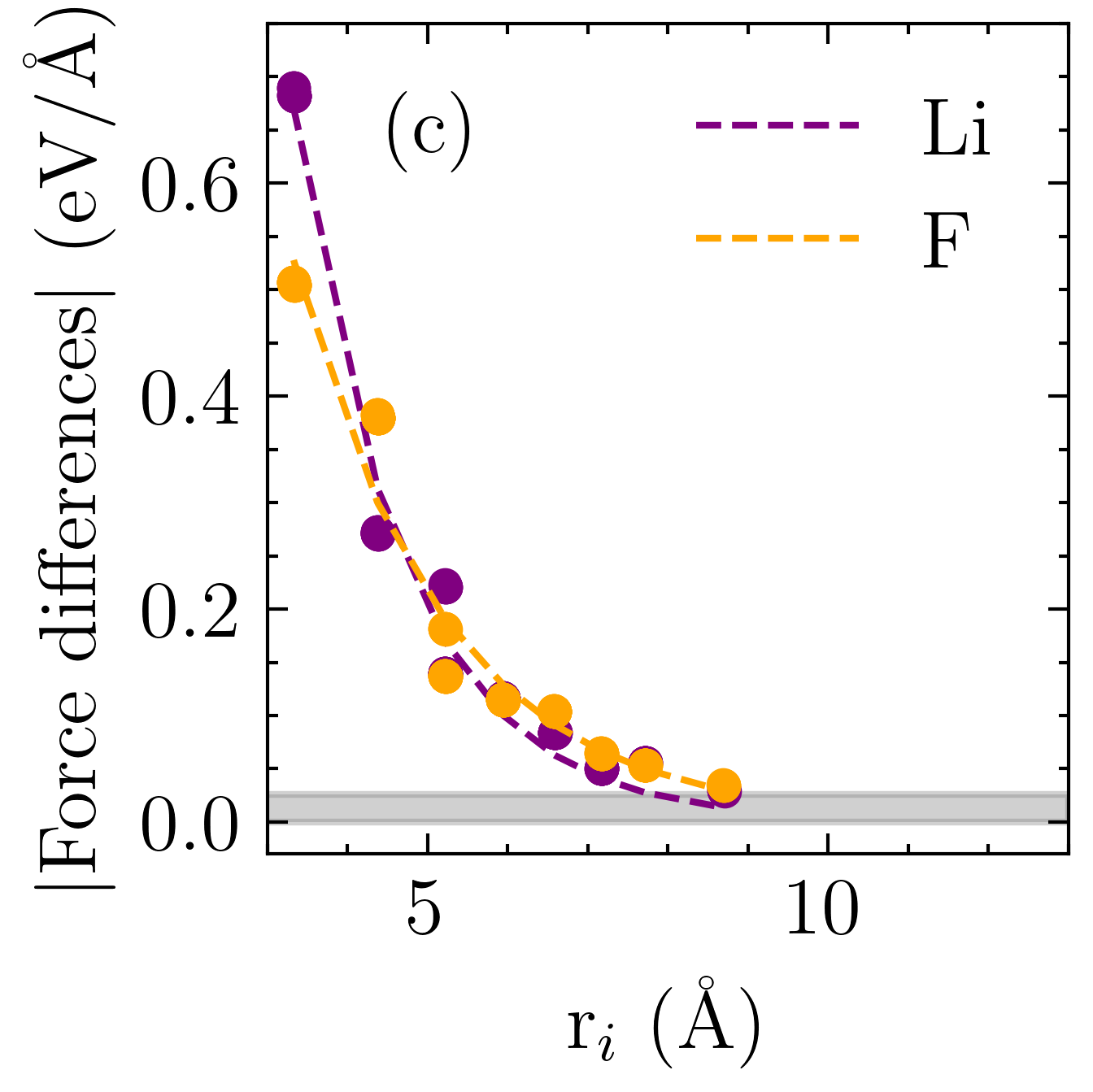}
\label{fig:lif3}
\end{subfigure}
\begin{subfigure}{4.2cm} \includegraphics[width=1.0\linewidth]{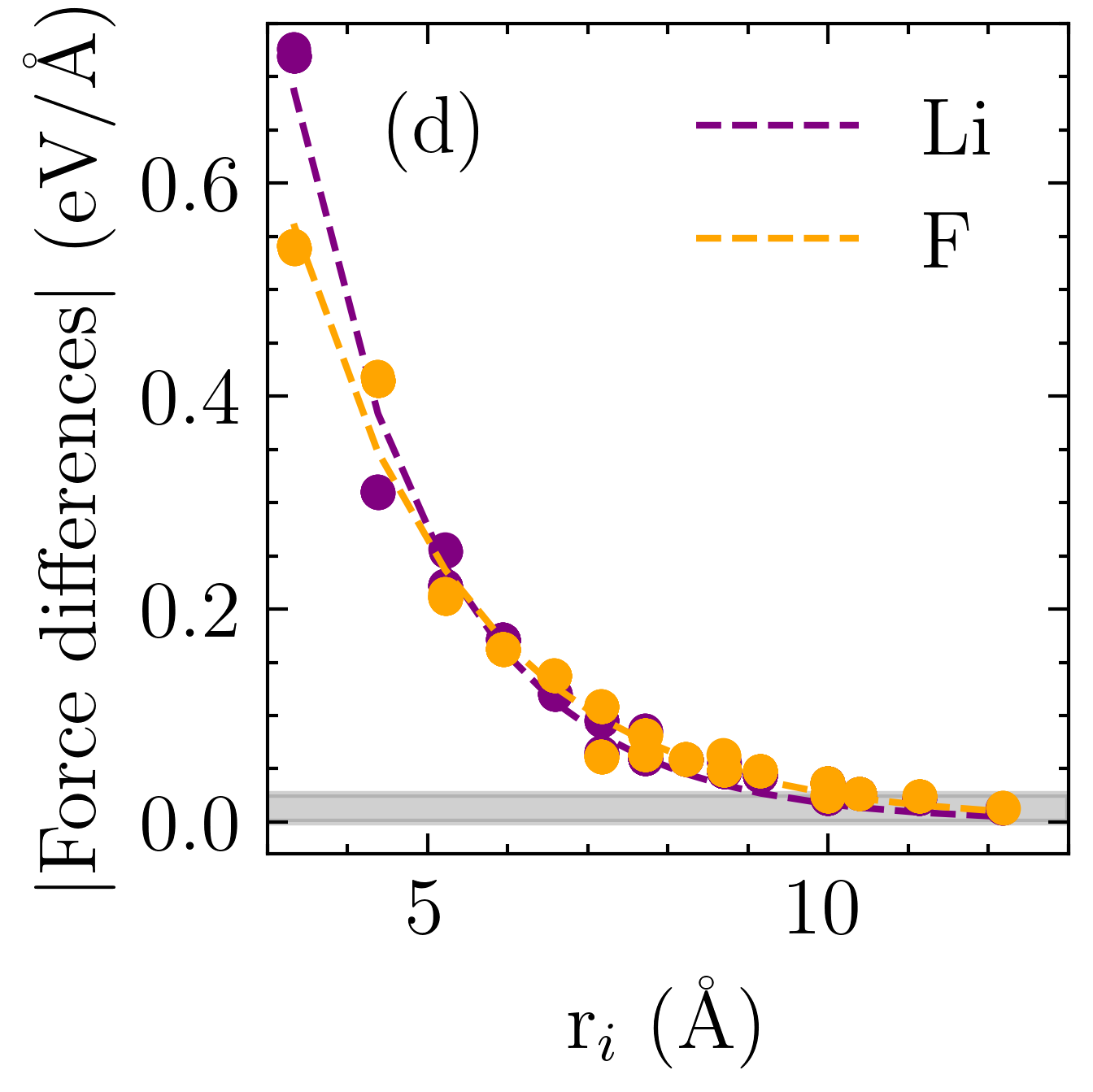}
\label{fig:lif4}
\end{subfigure}
\caption{Muon-induced forces on host atoms with respect to their distance from the muon $r_{i}$ in the Mu$^+$ charged \emph{unrelaxed} supercell calculation. Dashed  lines are fits  (see text) for each atomic species (symbols of the same color). Closest (bound) atoms are not shown. LaCoPO in a 3$\times$3$\times$2 supercell: (a) calculated forces with the muon; (b) force differences with and without the muon.  LiF  forces differences in a (c) 3$\times$3$\times$3 supercell;  (d) 4$\times$4$\times$4 supercell. In all panels, the maximum of the horizontal gray bar indicates the threshold on force differences $\Delta F$.}
\label{fig:lacopo_lif_scfconv}
\end{figure}

\begin{figure}
\centering
\includegraphics[width=1.0\linewidth]{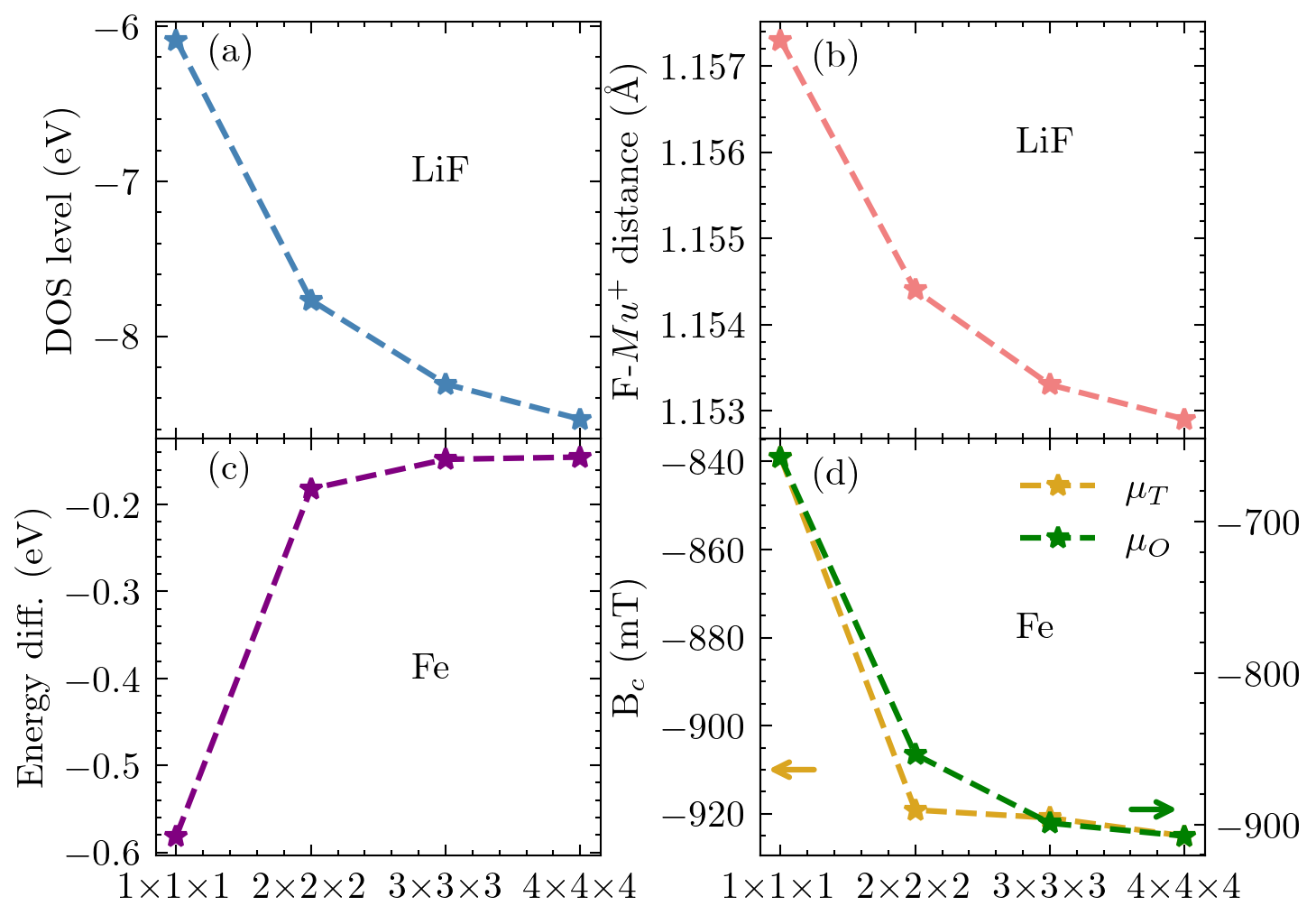}
\caption{ Convergence of muon-related  \emph{relaxed} calculations vs. supercell size. LiF: (a) position of the Mu$^+$ DOS peak with respect to the Fermi energy, and (b) F--Mu$^+$ distance. Fe: (c) Total energy difference between \muT~and \muO, and (d) contact hyperfine field ($B_{\mathrm c}$).}
\label{fig:lifconvres}
\end{figure}

Fig.~\ref{fig:conv}  shows the flowchart describing the supercell convergence algorithm. Starting from the input structure, the first step (a) consists of generating a nearly cubic supercell (implemented using the \texttt{CubicSupercellTransformation} Python class of the Pymatgen package~\cite{ONG2013314}). The size of the generated supercell is controlled by two parameters:  a minimum length for the supercell and a minimum number of atoms allowed.  These are optional input parameters, whose default values are $l_{at}+1~\text{\AA}$ and $N_{at}+1$, respectively, where $l_{at}$ is the length of the smallest lattice vector of the input structure and $N_{at}$ is the number of atoms in the input structure. Step (b) is accomplished by selecting one Voronoi interstitial node in the unit cell by means of the \texttt{VoronoiInterstitialGenerator} Python class of the Pymatgen package, and inserting an atomic site at that position in the supercell using a hydrogen pseudopotential, in order to mimic the muon \footnote{The function \texttt{VoronoiInterstitialGenerator} also implements an internal logic to cluster Voronoi nodes close to each other and remove nodes too close to the atoms of the hosting system.}. The forces acting on all atoms are then obtained from a converged self-consistent DFT calculation in step (c) and are used to check for convergence in step (d). 

Figure \ref{fig:lacopo_lif_scfconv} illustrates, both for LaCoPO (a metal) and for LiF (an insulator), that the forces obtained with a single SCF calculation decay exponentially with their distance from the muon. The decay length $\lambda$ is obtained as the best fit to $F \exp(-\lambda r_{i})$. Notice that an unrelaxed charged supercell, even without the muon, can show forces on the host atoms. For this reason we always consider the difference between the force on each atom with and without the muon (in the uncharged case the latter vanish). For example, forces on La atoms in unrelaxed charged LaCoPO, Fig.~\ref{fig:lacopo_lif_scfconv}(a), decay to a constant, representing the effect of (spurious) electronic charge doping, which is correctly subtracted when force differences are considered in Fig.~\ref{fig:lacopo_lif_scfconv}(b).

We assume that convergence is reached when atomic forces decay below a given threshold $\Delta F$, which in the workflow is an optional input parameter, with the default set to $1 \times 10^{-3}$ Ry/Bohr or 0.0257 eV/\AA \footnote{Atoms where the force may vanish due to PBC are excluded from the fit.}. To obtain a converged supercell, two conditions that ensure vanishing forces within the cell have to be satisfied: the minimum host atomic force is less than $\Delta F$ and the maximum $r_\mathrm{{i}}$  distance is greater than the minimum convergence distance, $\ln\frac{\Delta F}{F}/(-\lambda)$. If convergence is achieved, the workflow returns the supercell used in the last step and the corresponding transformation matrix with respect to the input structure. If convergence is not achieved, a larger supercell is generated and the loop goes back to step (c), provided that the maximum number of iteration is not exceeded.

In the case of LiF, Fig.~\ref{fig:lacopo_lif_scfconv}(c, d), convergence is achieved in a 4$\times$4$\times$4 supercell, Fig.~\ref{fig:lacopo_lif_scfconv}(d), using the default value of $\Delta F$ in the workflow, while forces are still above that threshold at the boundaries of the 3$\times$3$\times$3 supercell, as visible in Fig.~\ref{fig:lacopo_lif_scfconv}(c).

In order to verify our assumption that residual forces on atoms are a good proxy for convergence, we check explicitly the convergence of relevant muon-related quantities in LiF and Fe. At variance with the previous approach, here we relax all atomic positions for each supercell size, which becomes quickly very expensive. In LiF, we compute two quantities for four supercell sizes: the Mu$^+$ DOS level, i.e., the energy position from the Fermi energy of the Mu$^+$ peak in the density of states (DOS), that we show in Fig.~\ref{fig:lifconvres}(a), and the F--Mu$^+$ distance, shown in Fig.~\ref{fig:lifconvres}(b), for the muon at the F--Mu$^+$--F site~\cite{PhysRevB.87.121108,moeller2013}. 
As shown, they meet indeed the convergence criterion in the same 4$\times$4$\times$4 supercell found by the much more efficient force-difference method.
We run similar checks for Fe, where  the two candidate muon sites are the octahedral (\muO) and tetrahedral (\muT) interstitial sites. Here, we compare and show in Fig.~\ref{fig:lifconvres}(c-d) the total energy difference between \muO~and \muT, and the calculated contact field at both sites. The force difference method predicts the supercell convergence to be achieved with a 3x3x3 conventional cubic cell, which is indeed confirmed by the trend of the quantities shown in Figure~\ref{fig:lifconvres}(c-d).

\subsubsection{Unique sites selection}
\label{sec:clusanaly}
For each starting position in the above protocol, the relaxed candidate site eventually reaches a minimum of the potential energy surface. Despite the fact that the set of initial points is reduced by symmetry, symmetry-equivalent equilibrium positions are still often produced. These correspond, within numerical noise, to either the same site or to symmetry related sites. The conditions that they must meet to be recognized as equivalent is to have the same total energy within a given tolerance energy $\Delta{\varepsilon}$, and to be closer than a given tolerance distance $\Delta d$ (either directly, or after a suitable symmetry operation). 

In order to identify symmetry-inequivalent muon sites, clustering algorithms~\cite{ceriottireview} based on the conventional hierarchical and k-means method have been proposed~\cite{liborio2018}. However, here we introduce a more tailored, simpler and efficient method based on three quantities: (i) the Euclidean distance between sites, (ii) the total energy differences, and (iii) crystallographic and magnetic symmetries. 
The usage of these quantities are controlled by the tolerance values, namely: $\Delta{\varepsilon}$, $\Delta d$, and a separate tolerance distance $\Delta{s}$ for symmetry related replicas. Our clustering algorithm considers only the symmetry of the muon site, since equivalent muon sites will induce equivalent crystal distortions. The algorithm is initialized with the unit cell, with its symmetry operations and the magnetic unit cell (if applicable). All relaxed muon positions and their corresponding total energies are then added. Unique candidate muon sites are selected according to the following three steps:

\begin{enumerate}
\item{Remove duplicate positions within intersite distance $\Delta d$ and energy $\Delta{\varepsilon}$ tolerance, while retaining only that with the lowest energy.}
\item{Remove the crystal-symmetry equivalent positions within symmetry $\Delta s$ and energy $\Delta{\varepsilon}$ tolerance, while retaining only the lowest-energy inequivalent ones.}

\item{Add magnetically inequivalent positions. This is performed generating the replica of the crystalline-equivalent sites in the magnetic unit cell and removing those that are equivalent under magnetic symmetry. Any new site generated this way is resubmitted to the standard workflow in step III of  the ``Automated workflow design'' section.} 
\end{enumerate}
Step 1 saves a list of $n_2\leq N_\mu$ positions and their corresponding energies, step 2 adds a restricted list of $n_3<n_2$  distinct candidate muon sites with their energies and step 3 (when applicable) adds a separate list of $m$ magnetically inequivalent site(s). Finally, the algorithm outputs $n_3$ unique candidate muon sites and, in case, the new $m$ magnetic inequivalent positions whose structural relaxations must be performed. The algorithm is based on the Pymatgen libraries that can reproduce the symmetry operations pertinent to the input structures of the workflow. The default tolerance values are $\Delta{d} = 0.5$ \AA, $\Delta{s}= 0.05$ \AA, $\Delta{\varepsilon} = 0.05$ eV. 

\begin{figure}
\centering
\centering
\begin{subfigure}{3.6cm}
\includegraphics[width=\textwidth]{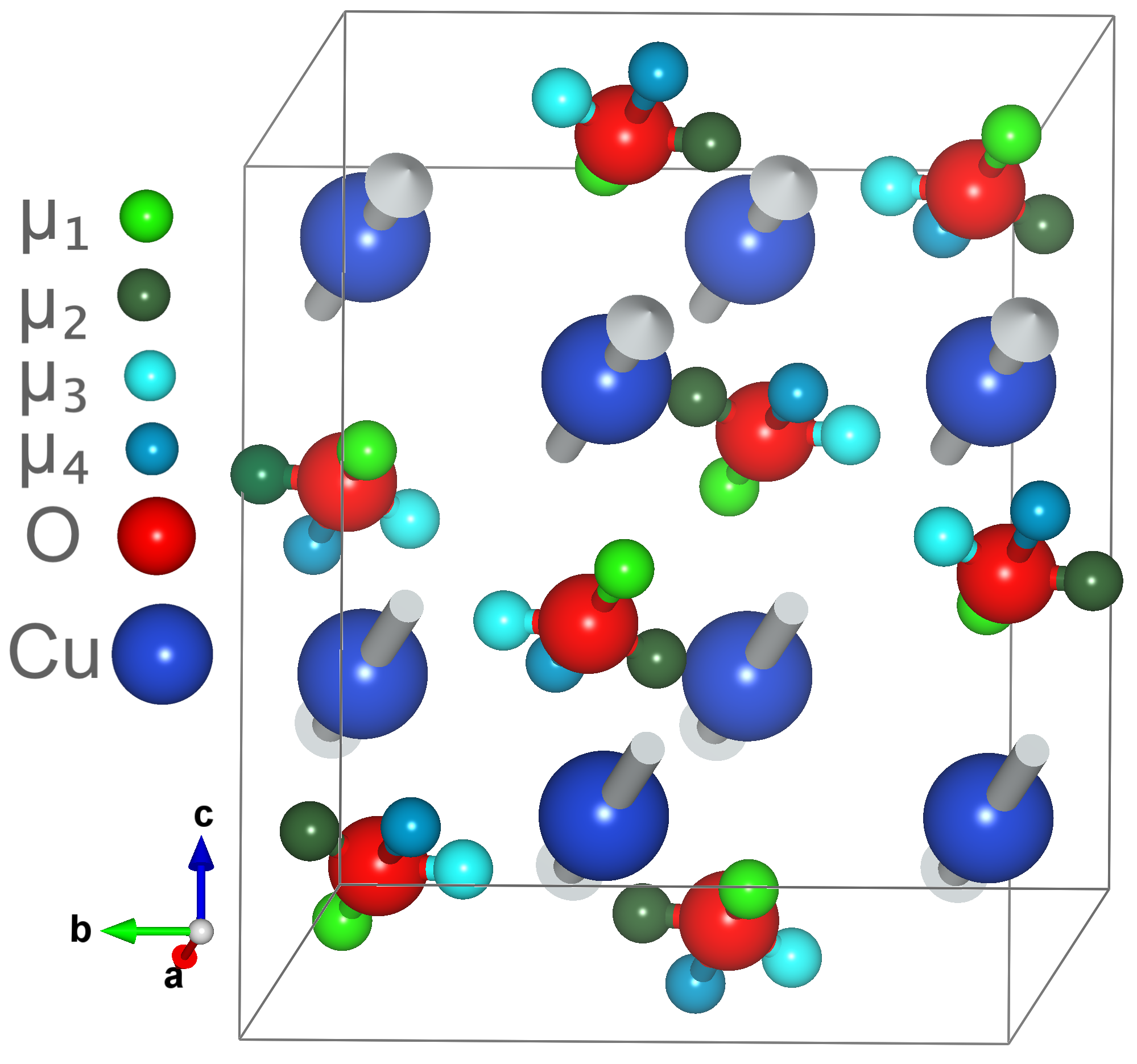}
\caption{}\label{fig:cuomusites}
\end{subfigure}
\begin{subfigure}{4.5cm}
\includegraphics[width=\textwidth]{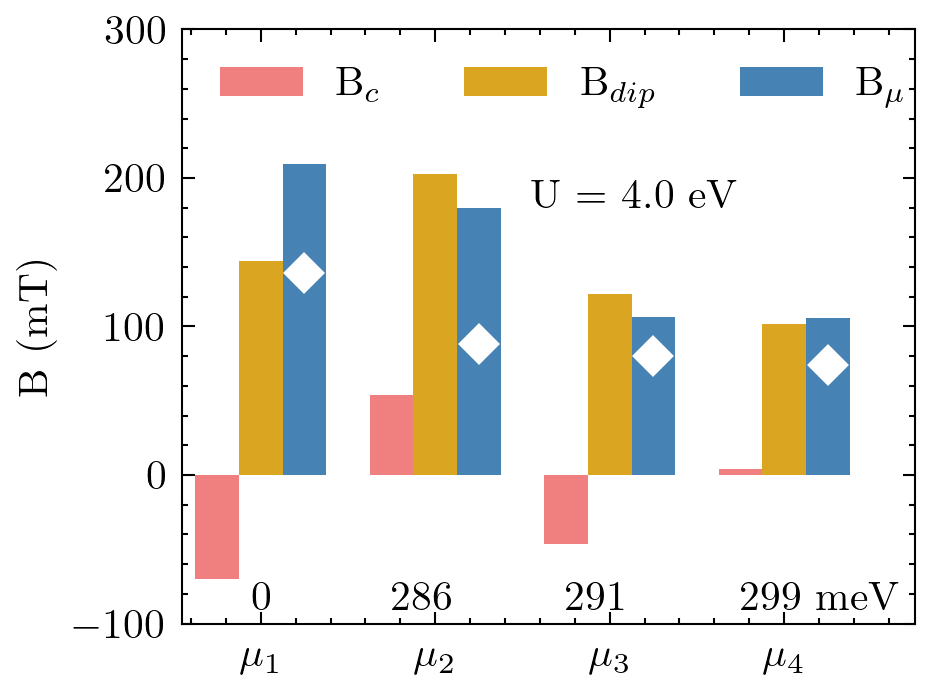}
\caption{}\label{fig:CuObestU}
\end{subfigure}
\begin{subfigure}{4.5cm}
\includegraphics[width=\textwidth]{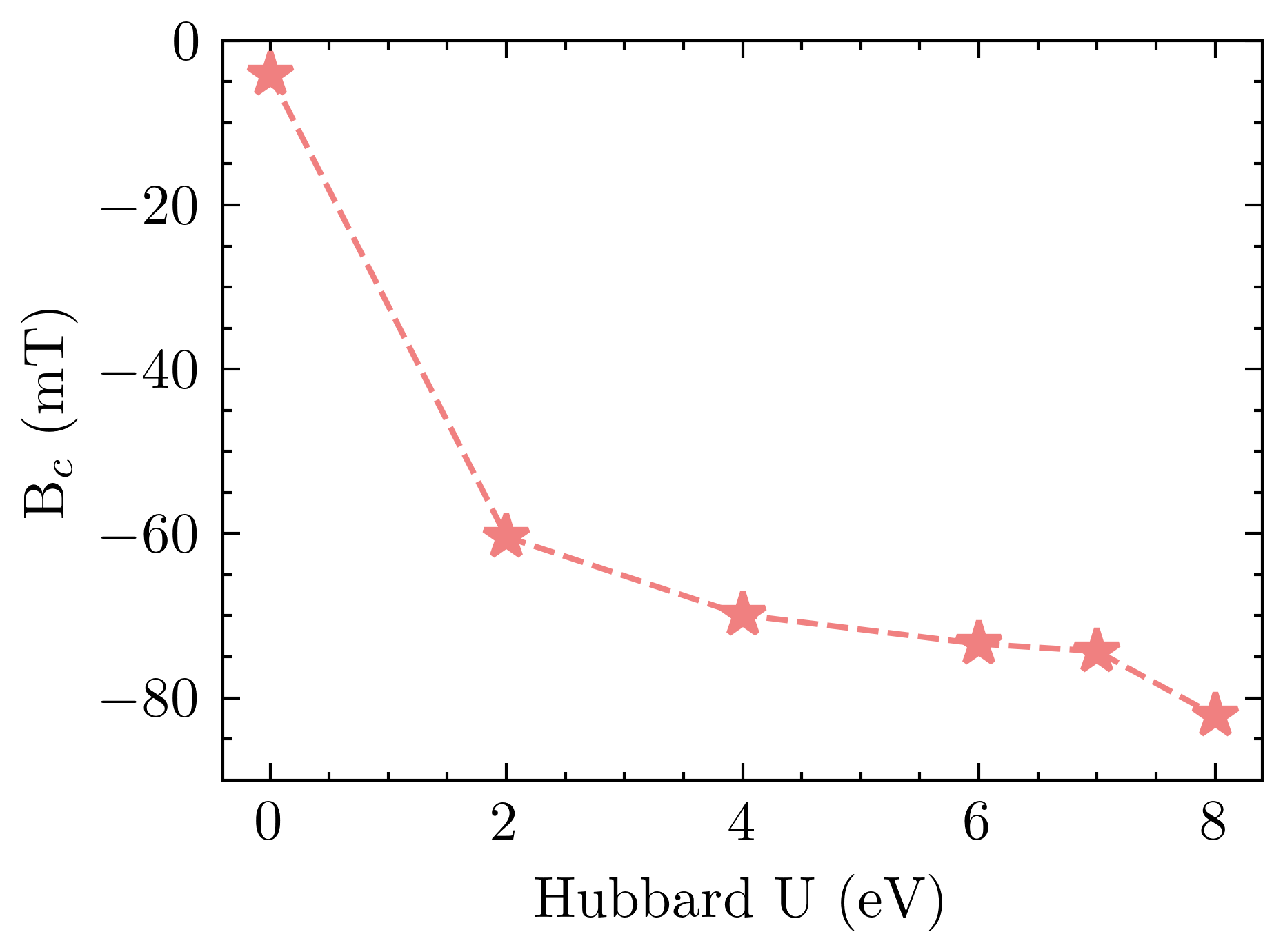}
\caption{}\label{fig:cuouvalues}
\end{subfigure}
\caption{ (a)  The CuO magnetic unit cell with inequivalent muon sites ($\mu_1-\mu_4$);   
(b) internal field contributions for $\mu_1-\mu_4$: dipolar, $B_\mathrm{dip}$,  contact, $B_\mathrm{c}$ and  total, $B_\mu$  compared to experimental values (diamonds). (c) Contact contribution  $B_\mathrm{c}$ to the internal field for the $\mu_1$ site (light-green) for different values of the Hubbard U correction.
}
\label{fig:ufield}
\end{figure}

\subsubsection{Electronic correlations: DFT+U}
A large fraction of the materials investigated with \musr{} present magnetic ground states and are possibly characterized by large electronic correlations. This  notoriously poses a challenge for DFT, due to the known tendency of the local density (LDA) and generalized gradient approximations (GGA) of the exchange and correlation contribution to over-delocalize the valence electrons, which is the main reason behind the poor description of the electronic and magnetic properties in strongly correlated electron systems. In turn, this may affect the prediction of the muon sites in magnetic compounds and most certainly that of their hyperfine interactions.

A number of advanced corrections already exist to amend this problem. These include the usage of hybrid functionals, such as the Heyd–Scuseria–Ernzerhof (HSE) functional\cite{10.10631.1564060}, the meta-GGA exchange-correlation functionals, in particular the strongly constrained and appropriately normed (SCAN) approach~\cite{scan2015} and its more recent variants such as rSCAN \cite{Bartok_rscan_2019} and r2SCAN \cite{Furness_r2scan_2020}, and the GW method~\cite{Aryasetiawan1998}. However these approaches, even though semi- or non-empirical (i.e., not parameter dependent), are very computational demanding~\cite{bonacci2023}.
An alternative route is provided by the DFT+U method~\cite{anisimov1991,anisimov1997, cococcioni2005}, which delivers an excellent trade-off between accuracy and efficiency. 
However, the major drawback for automation is that the Hubbard U value is system dependent. One option to obtain its value is using the linear response approach~\cite{cococcioni2005,kulik2006,timrov2018}, that however significantly increases the calculation cost. 
Nevertheless, even though the values of U vary with each compound, it has been shown that adapting a set of optimal fixed calibrated U values provides an acceptable compromise~\cite{Uvalues2006} for improving the description of electronic correlations and further facilitates the seamless automation of high-throughput workflows \cite{Torelli_2019,chen2019,doi:10.1021/acs.jpcc.2c06733,torelli2020,C9NA00588A, doi:10.34133/2022/9857631,matprojvalues2020, horton2019}. In the current version of the DFT+$\mu$ workflow we have therefore implemented as default the optimal Hubbard U values reported in Ref.~\citenum{Uvalues2006} for the following transition-metal ions: Co, Cr, Fe, Mn, Mo, Ni, V, W, Cu, which were obtained from the calibration of formation energies in transition-metal oxides. We note that, alternatively, the user may also provide a set of custom Hubbard U values as an input to the workflow.

As an example, let us consider cupric oxide (CuO).
Its ground state may be seen as a chain antiferromagnet (AFM), ordering three-dimensionally in an A-type collinear AFM structure. It is well known that LDA or GGA incorrectly yields a metallic and non-magnetic ground state for CuO\cite{PhysRevB.87.115111}, whereas DFT+U \cite{PhysRevB.90.014437,PhysRevB.73.235206, PhysRevLett.95.086405} reproduces the known antiferromagnetic ground state.  
Zero-field \musr{} (ZF-\musr{}) detects four distinct internal fields (74 mT, 80 mT, 88 mT, 136 mT) at low temperatures \cite{PhysRevB.38.2836,duginov1994,DAWSON19971383,nishiyama2001} and
four distinct field values are indeed reproduced with U = 4.0 eV, the adopted reference value for Cu~\cite{Uvalues2006}. They correspond to the two pairs of crystallographic magnetically inequivalent sites bound to each O, shown in Fig.~\ref{fig:ufield}(a). Their dipolar $B_{\mathrm{dip}}$ and contact $B_{\mathrm{c}}$ fields, as well as their vector sum modulus $B_\mu$ are compared with the experiment in Fig.~\ref{fig:ufield}(b). The agreement is reasonably good and Fig.~\ref{fig:ufield}(c) shows the dependence of the contact term $B_c$ on the value of the Hubbard U. While a vanishing value is obtained for $\text{U} = 0$~eV, finite values of U all provide reasonable predictions of $B_c$. This example indicates that the muon hyperfine interaction is sensitive to the treatment of the electronic correlations, and highlights how not including a finite U value results in an incorrect prediction also at the qualitative level.

We note that the positive muon stabilizes in the minimum of the electrostatic potential, more likely being attracted to the most electronegative atomic sites, implying that implantation site is determined by the larger scale of the electric interactions and it may be not sensitive to the much smaller scale of  magnetic interaction. This assumption is borne out by the CuO case, where the site is also correctly predicted with U = 0 eV, whereas the correct self consistent hyperfine coupling requires the Hubbard correction. However, the use of DFT+U+$\mu$ is known to be relevant also for the localization problem in other cases, such as that of MnO \cite{bonfa2024}.

\subsection{Validation cases}
\label{sec:validation}
In this section we present a selected set of cases to validate the \texttt{FindMuonWorkChain}. The set includes non-magnetic and magnetic metals and magnetic insulators. To demonstrate the level of automation possible, thanks to our workflows, the calculations reported below were executed providing only the minimum required input to the workflow, i.e., the crystal and magnetic structures, and keeping the default settings for other optional inputs; furthermore, the \texttt{IsolatedImpurityWorkChain} is used to obtain the converged supercell size for each case. 

\begin{figure}[tb] 
\centering
\includegraphics[width=6.2cm]{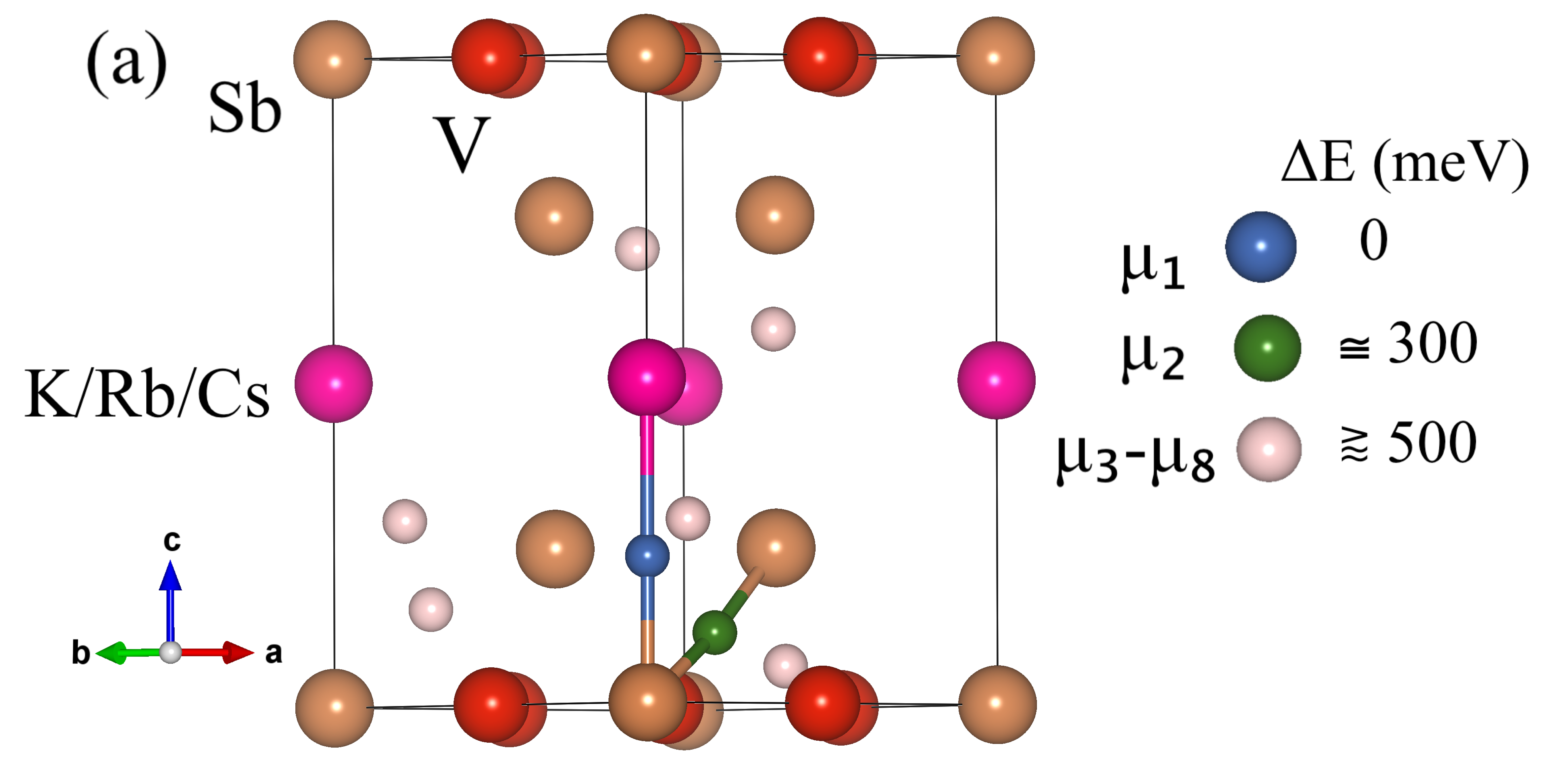}
\includegraphics[width=6cm]{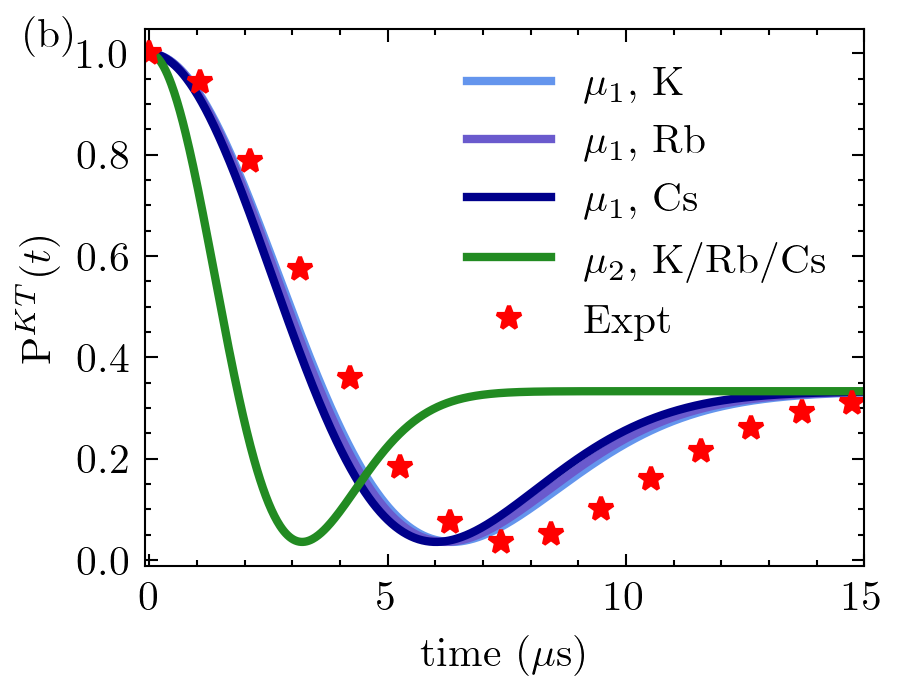}
\caption{(a) AV$_3$Sb$_5$ (A= K, Rb, Cs) unit cell with eight candidate muon sites, labelled $\mu_1$ to $\mu_8$. (b) Comparison between Kubo--Toyabe polarization function, Eq.~\eqref{eq:KT}, for the $\mu_1$ and $\mu_2$ sites, and the experimental result~\cite{guguchia_tunable_2023}. }
\label{fig:RbV3Sb5}
\end{figure}

\subsubsection{ ZF polarization in the Kagome metals, \texorpdfstring{AV$_3$Sb$_5$}{Lg} (A = K, Rb, Cs)}
KV$_3$Sb$_5$, RbV$_3$Sb$_5$ and CsV$_3$Sb$_5$ are Kagome superconductors where charge density wave orders and so-called time reversal symmetry breaking (TRSB) phases
have been extensively characterized by $\mu$SR spectroscopy in the past few years~\cite{kagome_charge_order_2022,guguchia_tunable_2023}. Here, using the \texttt{FindMuonWorkChain}, we compute the muon stopping sites in these samples with the hexagonal lattice (P6/mmm space group, No. 191), and we further simulate the time-dependent muon ZF polarization spectra.

The supercell workflow produces identical transformation matrices in all three systems:
$$
T=
\begin{bmatrix}
    -2 & -2 & 0\\ 
    -1 & 1 & 0\\
    0 & 0 & -1
\end{bmatrix}.
$$
The default value of the muon spacing identifies 30 initial positions for the calculations. The resulting candidate muon sites, shown in  Fig.~\ref{fig:RbV3Sb5}(a), are the same for the three compounds. The minimal assumption mentioned earlier is that the lowest-energy site, labelled $\mu_1$ (blue atom in Fig.~\ref{fig:RbV3Sb5}(a)) is occupied, while the remaining higher energy sites ($\mu_2-\mu_8$, green and pink atoms) are not. Site $\mu_1$ is located between the  K/Rb/Cs and Sb ions (at distance 2.60/2.73/2.72 \AA{} and 1.78/1.77/1.75 \AA{}, respectively) along the $c-$axis, while site $\mu_2$ bonds to two Sb ions at 2.2 \AA{} and two V ions at 1.8 \AA. In Fig.~\ref{fig:RbV3Sb5}(b), we show the plot of the Kubo--Toyabe depolarization function~\cite{kubo_toyabe,kagome_charge_order_2022} at sites $\mu_1$ and $\mu_2$, which reflects the field distribution at the muon site due to the nuclear moments, with the second moment (width) of the field distribution computed in the limit of strong quadrupolar interactions~\cite{PhysRevB.20.850, second_moment2021}, taking into account isotope averages and the lattice distortions induced by the muon. Despite the semi-classical approximation, the polarization function for site $\mu_1$ shown in Fig.~\ref{fig:RbV3Sb5}(b) shows the best agreement with the experimentally measured asymmetry in Ref.~\citenum{guguchia_tunable_2023}, thus validating the predicted site. Better agreement with experiment can be obtained incorporating the full quantum treatment of the muon-nuclear interactions including effects of the electric field gradient.

\begin{figure}[tb]
\centering
\includegraphics[width=8cm]{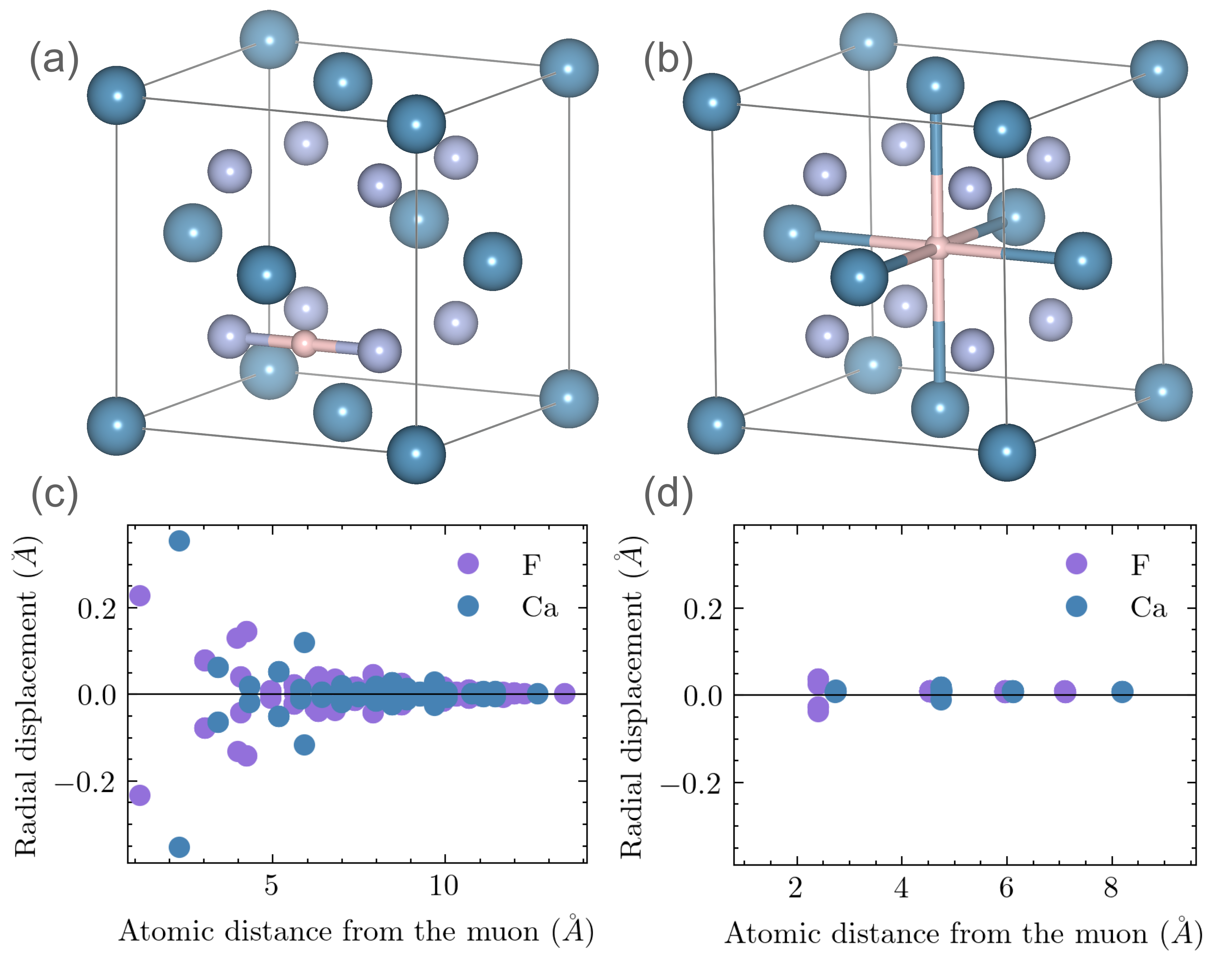}
\caption{Muon sites (pink atoms) in CaF$_2$: (a) diamagnetic state (Mu$^+$) and (b) paramagnetic state(\Mudot). Muon-induced displacements of the host atoms from equilibrium for Mu$^+$ (c) and  for \Mudot~(d).}
\label{fig:caf2}
\end{figure}
\subsubsection{Muon charge states in CaF\texorpdfstring{$_{2}$}{Lg}}

In order to  benchmark the \verb|FindMuonWorkchain| for handling Mu$^+$  and \Mudot~calculations, we select calcium fluoride (CaF$_2$), known to give rise both to the well characterized\cite{PhysRevB.33.7813} F-Mu$^+$-F center, and  to a \Mudot~state localized at the centre of the primitive cell~\cite{PhysRevB.87.121108,moeller2013}.    

The charge state is controlled in the workflow by the Boolean input parameter \verb|charged_supercell|, which is \verb|True| by default, yielding the Mu$^+$ state, and must be set to \verb|False| for the \Mudot~state. The converged supercell size is 3$\times$3$\times$3  for the charged state and 2$\times$2$\times$2 for the neutral state. This is a general finding: larger supercell sizes are required in the presence of charged interstitials with respect to the neutral case due to the longer-range Coulomb interaction between spurious periodic images.

For the charged state, the lowest energy sites consists of the muon bonding linearly to two neighbour F nuclei with distance 1.13 \AA{}, forming the so called F--Mu$^+$--F bond and shown in Fig.~\ref{fig:caf2}(a). This site is in agreement with earlier findings~\cite{PhysRevB.87.121108,moeller2013}. On the other hand,  the lowest energy site in the neutral supercell calculations   consists of a muon at the centre of the unit cell, \footnote{Here, we point out that the default spacing (1~\AA) for generating the muon initial positions gives only 4 initial positions. Reducing this spacing to 0.8~\AA{} (13 initial positions) also gives the same lowest energy muon sites as reported, but it generates also a higher energy site not found by the former grid.} as shown in Fig.~\ref{fig:caf2}(b).
Figures \ref{fig:caf2}(c) and (d) show the displacements from equilibrium of the host Ca and F ions vs.~their distance from the muon for the lowest energy Mu$^+$ and \Mudot~sites, respectively.  As expected, much larger displacements are obtained close to charged Mu$^+$ than to neutral \Mudot.   

\begin{figure} [ht]
\centering
\includegraphics[width=8cm]{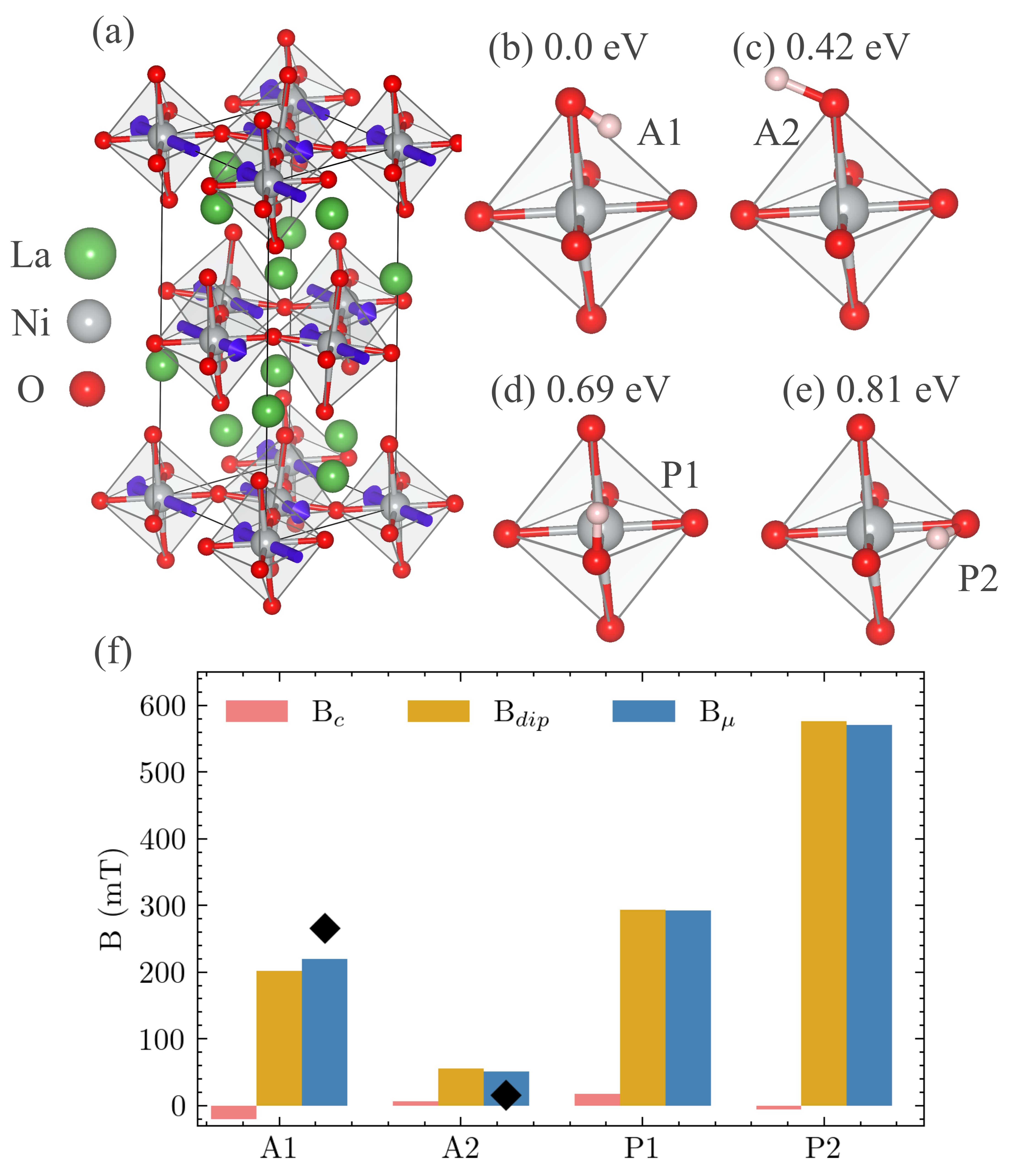}
\caption{(a) The magnetic unit cell of La$_2$NiO$_4$. (b-e) Muon positions (pink spheres)  bound to oxygen (labelled A for apical and P for planar) and their relative DFT energies in eV. (f) The contributions to the internal field at the muon in the four sites, where the black diamond symbols indicate reference experimental values.}
\label{fig:la2nio4}
\end{figure}

\subsubsection{Muon internal field in \texorpdfstring{La$_2$NiO$_4$}{Lg}}
We further demonstrate the role of electronic correlations in the search of the muon sites in La$_2$NiO$_4$, an antiferromagnetic insulator with the layered perovskite structure~\cite{Rodriguez_1991} shown in Fig.~\ref{fig:la2nio4}(a), where ZF \musr{} measurements established the presence of long-range antiferromagnetic order \cite{MARTINEZ1992941,chow1996,chow1997,jestadt1999} by detecting two distinct muon spin precessions at internal fields of 14.8 mT and 265.6 mT. The DFT calculations are carried out in the $2 \times 2 \times 1$ charged supercell given by the convergence algorithm, using 52 initial muon positions and setting the Boolean \verb|hubbard| input parameter to \verb|True| to use the default Hubbard $U=6.4$~eV for Ni. At variance with the literature, \cite{Lechermann_2022} the non-magnetic metallic state obtained at U = 0 is replaced by an insulating antiferromagnetic ground state with average magnetic moment of 1.59~$\mu_B$ on Ni, in agreement with neutron experiments (reporting 1.68~$\mu_B$ and  1.62~$\mu_B$, respectively\cite{Rodriguez_1991,PhysRevB.40.4463}). A small induced contribution at the planar O site is also observed. 
The workflow produces four distinct candidate muon sites forming bonds of length $\sim 1~$\AA{} to an O site, as shown in Fig.~\ref{fig:la2nio4}(b-e): two  bound to the apical O site, labelled A1 and A2, and two bound to the planar O site, labelled P1 and P2, plus an unstable, very high-energy site ($>$ 2.5 eV), located at an interstitial site between two consecutive perovskite layers (not shown). By the minimal-energy argument adopted before, we expect  muons to localize at the A1 site, and possibly at the A2 site. 

This site assignment is confirmed by the  muon internal fields calculated within the workflow. Fig.~\ref{fig:la2nio4}(f) shows the Fermi contact $B_{\mathrm c}$ and the magnetic dipole $B_{\mathrm{dip}}$ contributions to the internal field $B_\mu$ at all four candidate sites. In all the cases, the contribution from $B_{\mathrm {dip}}$ is dominant while those from  $B_{\mathrm c}$ are almost vanishing. Figure \ref{fig:la2nio4}(f) reveals that the values of $B_\mu$ computed at sites A1 and A2 are in reasonable agreement with the high (265.6 mT) and low (14.8 mT) internal fields observed in experiments\cite{MARTINEZ1992941,chow1996,chow1997}, also considering the present neglect in the workflow of the quantum zero-point vibrations. Notice that our site assignment partially contradicts earlier non ab-initio speculations  based on magnetic dipolar field analysis, \cite{MARTINEZ1992941,chow1996,chow1997} where the low internal field was correctly assigned to a muon bound to the apical oxygen, whereas the high internal field was attributed to muon localizing at the planar P2 site. Our findings are strongly supported by a very good reproduction of the magnetic interactions.

\subsubsection{LaCoPO: Case study for a ferromagnetic metal}
\label{sec:LaCoPO}
We finally validate the workflow in LaCoPO, a ferromagnetic metal in which neutron diffraction \cite{PhysRevB.77.224431} measures  0.3 $\mu_B/$Co. ZF \musr{} measurements indicate the existence of a single muon site\cite{PhysRevB.87.064401} with internal field of 58.75~mT and the corresponding interstitial muon stopping site has been previously identified. We have performed DFT+$\mu$+U calculations, with the default Hubbard value for Co, U = 3.3 eV. A converged 3$\times$3$\times$2  charged supercell (144 atoms) is obtained, while the default value of the muon spacing  gives 20 initial positions (supercells) for the calculations. 

\begin{figure}
\centering
\begin{subfigure}{3.2cm}
\includegraphics[width=\textwidth]{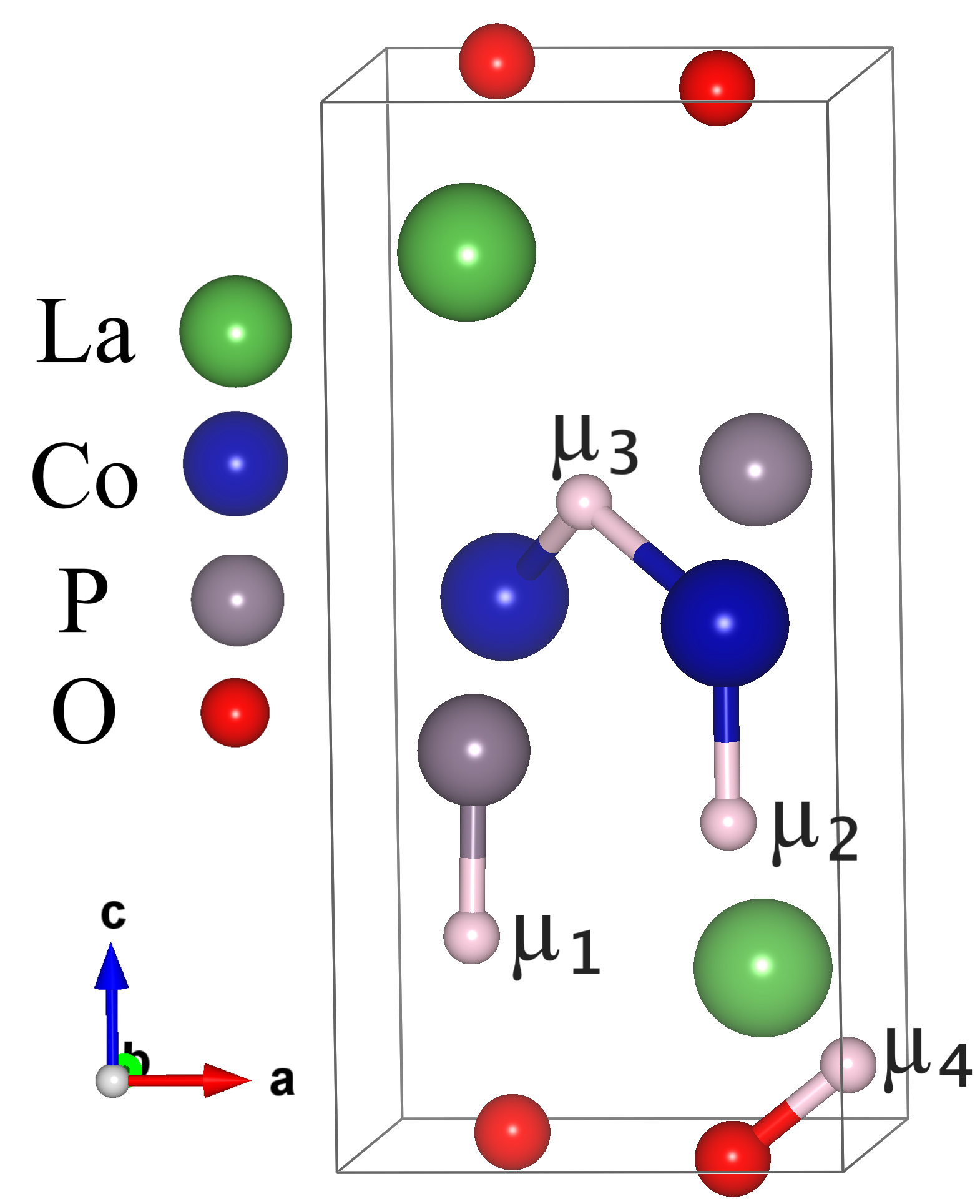}
\caption{}\label{fig:l1cop1}
\end{subfigure}
\begin{subfigure}{5.3cm}
\includegraphics[width=\textwidth]{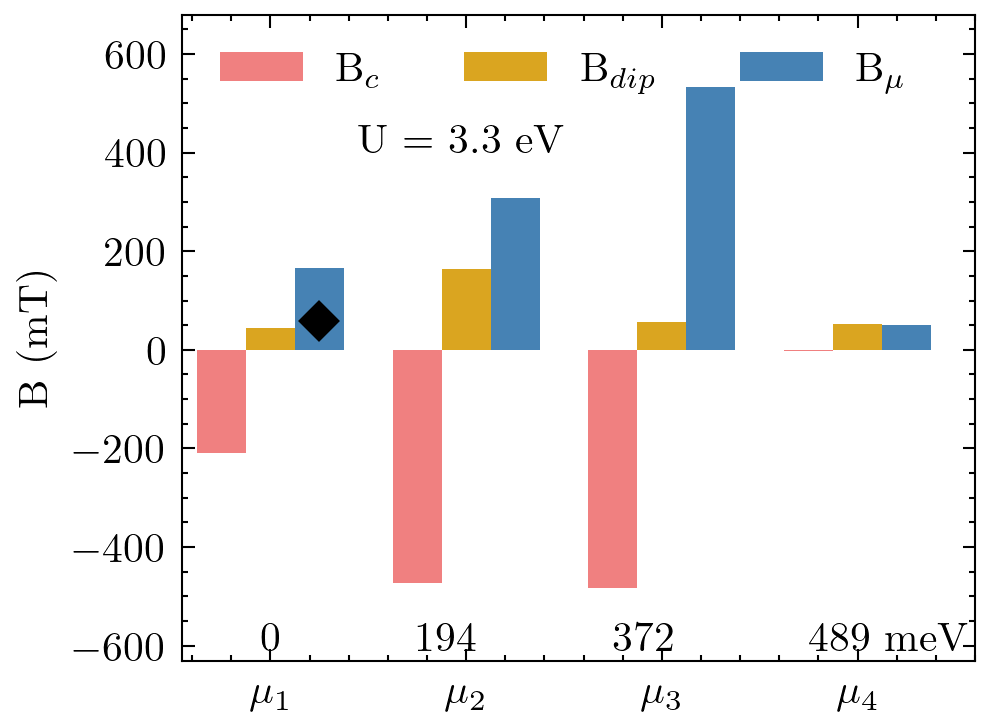}
\caption{}\label{fig:l1cop2}
\end{subfigure}
\caption{(a) Positions of four candidate muon sites (pink spheres) in the LaCoPO unit cell, labelled $\mu_1$ to $\mu_4$. (b) The Fermi contact $B_{\mathrm c}$ and dipolar $B_{\mathrm {dip}}$ contributions to the internal field $B_\mu$ at the four candidate muon sites computed using DFT+U. 
Negative  $B_{\mathrm c}$ values indicate opposite polarization to the bulk magnetization. The black diamond symbol indicates the experimental muon internal field. The labels above the horizontal axis indicate the total energy difference for each site with respect to $\mu_1$ in meV. }
\label{fig:lacopo}
\end{figure}

Figure \ref{fig:lacopo} (a) shows the LaCoPO unit cell and the positions of the four distinct candidate muon sites, labelled $\mu_1$ to $\mu_4$. The relative energies and the computed internal field contributions are shown  in  Fig.~\ref{fig:lacopo}(b). Site $\mu_1$, located 1.47 \AA{} away from the P site along the $c$ axis  (see  Fig.~\ref{fig:lacopo}) and  towards the La--O layer, is the lowest energy site, whereas the next lowest is site $\mu_2$, located  1.50 \AA{} away from the Co site  along the $c$ axis and towards the La--O layer. The assignment of the unique experimental internal field to site $\mu_1$ is in agreement with our minimal argument and with the earlier study~\cite{PhysRevB.87.064401}. Site $\mu_2$ has been reported to be unstable owing to the muon ZPM~\cite{PhysRevB.87.064401}. In Fig.~\ref{fig:l1cop2}, we show the contributions to the total internal field for all four candidate sites that we obtained. For all sites but $\mu_4$, a sizeable contribution due to the contact hyperfine term $B_{\mathrm c}$ is observed. Notice that the ferromagnetic ground state that we obtain shows an average magnetic moment localized on Co atoms of 0.7 $\mu_B$, more than twice the experimental value. The computed internal field $B_\mu$ at site $\mu_1$ is 166 mT. The value largely overestimates the experimental value, which is expected since $B_{\mathrm c}$ is directly proportional to the magnetic moment.

We emphasize that an improved agreement is already provided by the possibility to evaluate the distant dipole sums with the experimental magnetic structure, including the value of the moments.  However, users must be aware that the results provided by the workflow in this case are not ``self consistent'', since the contact contribution is obtained instead with the ab initio moments. A crude correction consists in re-scaling~\cite{onuorah2018} the value of $B_{\mathrm c}$ by a factor $m_{\mathrm{exp}}$/$m_{\mathrm{abinitio}}$, which yields $B_\mu$ = 46 mT, closer to the experimental value (note, however, that we do not expect a perfect quantitative agreement, also because we are not accounting for the muon ZPM). A better ab-initio solution would be to include a self-consistent correction of the magnetic moment using the reduced Stoner theory modification to the exchange-correlation functional~\cite{ortenzi2012,onuorah2018}.

\section{Future developments} 
One important feature that is still lacking in the current implementation of the workflow is the inclusion of anharmonicity and the muon ZPM effects~\cite{PhysRevLett.81.1873,bonfa2015,onuorah2019}. These corrections are known to influence both the stability of the muon sites and the estimation of the muon interactions, particularly for compounds where the energy barrier among a group of neighbor candidate sites is lower than their ZPE and may give rise to muon wavefunction delocalization (quantum diffusion). Future development in the workflow will optionally incorporate the muon ZPM  using the stochastic self-consistent harmonic approximation (SSCHA) method~\cite{onuorah2019}.

In the classical limit, thermal muon diffusion can be investigated using the nudged elastic band (NEB) method, that can track the barrier and the minimum energy path connecting the specified initial and final sites~\cite{neb1}. Future developments will optionally include this method in the workflow. This information contributes to the prediction of the temperature dependence of the \musr{} signal.

When the electronic contribution to the muon polarization function is negligible, numerical evaluation of the time evolution of the spin of the muon due to its interaction with the surrounding nuclei is straightforward~\cite{BONFA2021107719, PhysRevLett.129.097205, PhysRevLett.125.087201, PhysRevLett.56.2720}. Future development of the workflow will include the evaluation of dipolar and quadrupolar contributions~\cite{PhysRevLett.129.097205} thus allowing to extract accurate polarization functions from the first principles simulations.

Another improvement that may be considered is the implementation of the reduced Stoner theory correction to the exchange-correlation functional~\cite{ortenzi2012,onuorah2018} aimed at a more accurate description of the magnetic ground state and in turn the muon contact hyperfine contribution, particularly for itinerant electron systems where the magnetic moments are poorly estimated with DFT.

On the more technical side, while our automated selection of numerical parameters is able to deliver full automation and provides converged results, in some cases those values could be adapted, so as to run cheaper simulations without significantly affecting the physical results.
For example, in LaCoPO, the use of  a much smaller 2$\times$2$\times$1 cell (32 atoms, instead of the 144 atoms of the 3$\times$3$\times$2 cell obtained from the convergence workflow) and a larger  spacing of 1.2 \AA{} for the sampling muon grid (15 initial positions instead of 20), still gives identical results within numerical accuracy. Likewise, for LaNiO$_2$, the interstitial space can also be properly sampled by increasing the muon grid spacing to $d_\mu$=1.6 \AA{}, resulting in only 10 supercells to be computed instead of 52, thus saving computational cost. Therefore, some parameters of the workflow could be further optimized to minimize computational cost.

\section{Conclusion}
We have developed and validated an automated DFT-based workflow based on the AiiDA platform for the calculation of the muon stopping sites and hyperfine interactions in solids. 
The workflow has been successfully validated in a selected set of compounds including the Kagome structured superconductors, fluorides and magnetic insulating oxides. These validation cases  demonstrate the implementation of various tools for the analysis of \musr{} experiments.
The workflow has been highly optimized to require minimal input from the user, so that the crystal structure is the only mandatory information required to execute it. 
In particular, a fast algorithm for computing the convergence against the supercell size has been included in the workflow, and pre-defined values for Hubbard correction in DFT+U are provided in order to improve the treatment of strongly correlated systems.

Moreover, we provide predefined input settings of the workflow, and we demonstrate that these are sufficient to fully automate and provide accurate results for DFT+$\mu$ calculations. In addition, if computational resources are limited, it is possible for an experienced user to manually provide less optimal input parameters that still provide reasonably converged results in a shorter time, possibly with looser convergence thresholds.

Our automated workflow represents a powerful tool that will encourage, facilitate and promote the usage of ab-initio calculations by the \musr\ community, opening up the possibility to perform muon simulations routinely alongside experimental measurements.

\section{Methods: Computational Parameters}
For all calculations, we have used plane-wave based DFT as implemented in the Quantum ESPRESSO suite of codes \cite{qe2009}.
The standard solid-state pseudopotentials (SSSP) library set optimized for efficiency of the Perdew-Burke-Ernzerhof (PBE) functional (SSSP PBE efficency v1.3) have been used~\cite{prandini2018precision}. The validation cases reported above have been performed using default plane-wave and density cutoffs as provided by SSSP. Other test calculations (on LiF, Fe and CuO) that form part of the algorithm discussion have been performed using fixed plane-wave and density cutoffs of 60 Ry and 480 Ry, respectively. A Gaussian smearing with 0.01 Ry width and electronic convergence threshold of 10$^{-6}$ Ry were utilized for all calculations. The k-point grid distance of 0.301 \AA{} has been utilized to obtain a Monkhorst--Pack grid for Brillouin zone sampling, except for the supercell convergence algorithm validation reported in Fig.~\ref{fig:lifconvres}, where a grid distance of 0.1 \AA{} was used. The total computational resources required to run the examples in the validation section amounted to 342,580 CPU hours on the Leonardo Supercomputer (CINECA).

\section{Data availability}
The data of the results presented here are available on the Materials Cloud \cite{talirz_materials_2020} Archive at \href{https://doi.org/10.24435/materialscloud:yy-ds}{https://doi.org/10.24435/materialscloud:yy-ds}.

\section{Code availability}
\label{sec:codavail}
The \verb|FindMuonWorkchain| and the \verb|IsolatedImpurityWorkChain| are actively developed and can be downloaded from the respective GitHub repositories at \href{https://github.com/positivemuon/aiida-muon/}{https://github.com/positivemuon/aiida-muon/} and \href{https://github.com/positivemuon/aiida-impuritysupercellconv}{https://github.com/positivemuon/aiida-impuritysupercellconv}. These workflows rely on the \verb|aiida-quantumespresso| plugin and workflows available at \href{https://github.com/aiidateam/aiida-quantumespresso}{https://github.com/aiidateam/aiida-quantumespresso}. The code described in this paper is v.1.0.0 for both workchains.

\section{Acknowledgments}
IJO, MM, PB and RDR acknowledge financial support from the PNRR MUR project ECS-00000033-ECOSISTER. MB and GP acknowledge financial support from the NCCR MARVEL, a National Centre of Competence in Research, funded by the Swiss National Science Foundation (grant number 205602). This research was granted by University of Parma through
the action Bando di Ateneo 2023 per la ricerca. We acknowledge computing resources provided by the STFC scientific computing department's SCARF cluster and CINECA award under the IsCa4-SIEMTQM and CNHPC-1570115 projects. We acknowledge access to Piz Daint and Alps at the Swiss National Supercomputing Centre (CSCS) under PSI's share with project ID psi15 and psi18 and under MARVEL share with project ID mr32.

\bibliography{references}
\end{document}